\documentclass[conference,compsoc]{IEEEtran}

\usepackage[hyphens]{url}
\usepackage[numbers,sort&compress]{natbib}
\usepackage[breaklinks,colorlinks]{hyperref}
\usepackage[table,dvipsnames]{xcolor}
\hypersetup{citecolor=blue,linkcolor=blue}
\usepackage{amsmath,amsopn,amssymb}
\usepackage{bm}
\usepackage{subcaption}
\usepackage{endnotes,microtype,xspace,graphicx,fancyvrb,multirow}
\usepackage{booktabs}
\usepackage{array,underscore,relsize}
\usepackage[T1]{fontenc}
\usepackage{times}
\usepackage{fancyhdr,lastpage}
\usepackage{enumitem}
\usepackage[labelfont=bf,font=small,skip=5pt]{caption}
\usepackage{fp}
\usepackage{siunitx}

\usepackage[most]{tcolorbox}

\usepackage{balance}

\newcommand{\sys}{\mbox{\textsc{DualView}}\xspace}

\newcommand{\BL}[1]{\textcolor{blue}{}}
\newcommand{\JH}[1]{\textcolor{purple}{}}
\newcommand{\CDX}[1]{\textcolor{NavyBlue}{}}

\newcommand{\cc}[1]{\mbox{\smaller[0.5]\texttt{#1}}}
\newcommand{\eg}{e.g.,\xspace}
\newcommand{\ie}{i.e.,\xspace}
\newcommand{\llmterm}[1]{\mbox{#1}\xspace}
\newcommand{\TLLM}{\llmterm{T-LLM}}
\newcommand{\ULLM}{\llmterm{U-LLM}}

\newcommand{\AgentView}{AgentView\xspace}
\newcommand{\HumanView}{HumanView\xspace}
\newcommand{\AgentFileSystem}{Agent File System\xspace}
\newcommand{\HumanFileSystem}{Human File System\xspace}

\newcommand{\AgentShell}{AgentShell\xspace}
\newcommand{\HumanShell}{HumanShell\xspace}

\fvset{fontsize=\scriptsize,xleftmargin=8pt,numbers=left,numbersep=5pt}

\def\Snospace~{\S{}}

\newif\ifdraft\draftfalse
\newif\ifnotes\notesfalse
\ifdraft\else\notesfalse\fi

\input{glyphtounicode}
\newcolumntype{R}[1]{>{\raggedleft\let\newline\\\arraybackslash\hspace{0pt}}p{#1}}

\newcommand{\squishlist}{
\begin{itemize}[noitemsep,nolistsep]
  \setlength{\itemsep}{-0pt}
}
\newcommand{\squishend}{
  \end{itemize}
}

\usepackage{tikz}
\usetikzlibrary{positioning,arrows.meta,calc,fit,backgrounds}
\DeclareRobustCommand{\ClosedEyeIcon}{%
\raisebox{-0.7ex}{\includegraphics[height=2.4ex]{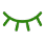}}}
\DeclareRobustCommand{\OpenEyeIcon}{%
\raisebox{-0.7ex}{\includegraphics[height=2.4ex]{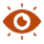}}}
\newcommand*\WC[1]{%
\begin{tikzpicture}[baseline=(C.base)]
\node[draw,circle,inner sep=0.2pt](C) {#1};
\end{tikzpicture}}

\newcommand*\CN[1]{\WC{#1}}

\definecolor{stepblue}{HTML}{2F6DB0}
\definecolor{stepred}{HTML}{A8423F}
\definecolor{steppurple}{HTML}{6E55A0}

\newcommand*\cA[1]{A-\CN{#1}}
\newcommand*\cB[1]{B-\CN{#1}}
\newcommand*\cC[1]{C-\CN{#1}}

\definecolor{viewgreen}{HTML}{2E9E5B}
\definecolor{vieworange}{HTML}{D9701E}
\newcommand*\StepFill[2]{%
\begin{tikzpicture}[baseline=(C.base)]%
\node[circle,fill=#1,inner sep=0.8pt,font=\bfseries\scriptsize](C){\textcolor{white}{#2}};%
\end{tikzpicture}}
\newcommand*\aN[1]{\StepFill{viewgreen}{#1}}
\newcommand*\hN[1]{\StepFill{vieworange}{#1}}

\usepackage{xstring}
\newcommand{\PP}[1]{
\vspace{2px}
\noindent{\bf #1.}
}

\newcommand{\goallabel}[1]{#1\xspace}
\newcommand{\gsec}{\goallabel{Security}}
\newcommand{\gutil}{\goallabel{Utility}}

\newcommand{\gdeploy}{\goallabel{Deployability}}

\newcommand{\reqlabel}[1]{#1\xspace}
\newcommand{\reqtrack}{\reqlabel{Track}}
\newcommand{\reqisolate}{\reqlabel{Isolate}}

\newcommand{\ra}[1]{\renewcommand{\arraystretch}{#1}}
\newcommand{\V}{\checkmark}
\newcommand{\X}{{\footnotesize $\times$}\xspace}

\newcommand{\boxbeg}{
\vspace{2px}
\noindent\begin{tabular}{|l|}\hline
\begin{minipage}{3.2in}
\vspace{2px}
\noindent
}

\newcommand{\boxend}{
\vspace{2px}
\end{minipage}\\ \hline
\end{tabular}
\vspace{-10pt}
}

\definecolor{principleaccent}{HTML}{1F4E79}
\definecolor{principlebg}{HTML}{F3F6FA}
\newtcolorbox{principlebox}[2]{%
  enhanced, breakable, sharp corners,
  colback=principlebg, colframe=principleaccent,
  boxrule=0pt, leftrule=2.5pt, toprule=0pt, rightrule=0pt, bottomrule=0pt,
  left=8pt, right=6pt, top=4pt, bottom=4pt,
  before skip=4pt, after skip=4pt,
  fontupper=\small,
  overlay unbroken and first={%
    \node[anchor=north west,
          fill=principleaccent, text=white,
          inner xsep=5pt, inner ysep=2pt,
          font=\bfseries\footnotesize]
         at ([xshift=2.5pt,yshift=-0pt]frame.north west)
         {#1~\textit{\textmd{(#2)}}};
  },
  before upper={\vspace{10pt}\par\noindent\ignorespaces},
}

\hypersetup{
  pdftitle={DualView: Preventing Indirect Prompt Injection in Personal AI Agents},
  pdfauthor={Juhee Kim, Woohyuk Choi, Taehyun Kang, Youngmin Kim, and Byoungyoung Lee}
}

\begin{document}

\title{\sys: Preventing Indirect Prompt Injection in Personal AI Agents}

\ifdefined\DRAFT
 \pagestyle{fancyplain}
 \lhead{Rev.~\therev}
 \rhead{\thedate}
 \cfoot{\thepage\ of \pageref{LastPage}}
\fi

\author{%
  \IEEEauthorblockN{Juhee Kim\textsuperscript{*}, Woohyuk Choi\textsuperscript{*}, Taehyun Kang, Youngmin Kim, and Byoungyoung Lee}
  \IEEEauthorblockA{Seoul National University}
}

\maketitle
\begingroup
\renewcommand{\thefootnote}{\fnsymbol{footnote}}
\footnotetext[1]{Equal contribution.}
\endgroup

\sloppy

\begin{abstract}
Personal AI agents that run on the user's local machine, such as
OpenClaw, automate daily tasks including web search, email, and file
management.
Their access to computer resources, including the network, file system,
and shell, exposes them to indirect prompt injection~(IPI) attacks.
Prior Dual LLM defenses block IPI by replacing untrusted data with
symbols that the agent can reference but not read.
However, they track untrusted data only inside the agent's context, so
when the agent saves and later rereads untrusted data, that data,
possibly an attacker's prompt, can return as trusted data rather than
as a symbol, which we call stored IPI.

Operating on the user's real environment, which humans and programs
share, is what makes agents like OpenClaw practical, and is exactly
why a defense that ignores it is incomplete.
Preserving symbols in such an environment is hard, because humans and
programs need original data.
We present \sys, which extends untrusted data tracking from the agent's
context to the user's environment, including the file system, shell,
network, and other agents, by giving each channel two views.
In \AgentView, the agent sees untrusted data as symbols even after
writing it out and reading it back, blocking stored IPI, while
\HumanView preserves original data for humans and tools.
\sys routes each tool call to the right view and synchronizes data
across the two views.
\sys deploys as an OpenClaw plugin using only tool hooks, without
changing the agent's tool-call logic or tool implementations.
\sys deterministically prevents instructions in untrusted data
from directly steering the agent's tool calls; this guarantee does not
depend on recognizing the evaluated attack templates.
In our evaluation on an IPI benchmark and PinchBench, \sys blocked
every tested IPI attack, including stored IPI. On PinchBench, its
utility was within 1.8 percentage points of the unprotected baseline
with Claude Haiku 4.5 and within 6.4 points with Claude Sonnet 4.6.
\end{abstract}

\section{Introduction}
\label{s:intro}

Personal AI agents are useful because they assist users by running in
the same computer environment where users already work.
Users often allow the agents to read local files, fetch external
data, run shell commands, send email, process webhook payloads, and
communicate with other
agents~\cite{react,toolformer,chatgpt-tools,microsoft-copilot,mcp}.
These capabilities let the agent not only answer questions but also
handle routine computer tasks for the user.

A representative example is OpenClaw~\cite{openclaw}, a local-first
AI agent that runs on the user's machine and exposes such tools.
A user can ask the agent to ``summarize market news and save notes to
\cc{./report.md}.'' To complete the task, the agent searches the web,
fetches web pages, and writes a local file.
The user's computer environment is also where humans and non-agent
programs continue to read, edit, and reuse the agent's results.

The same capabilities create the security problem.
When an agent reads external data and can act on the user's resources,
it is exposed to \emph{indirect prompt
injection}~\cite{greshake2023not,liu2024formalizing,injecagent}.
A remote attacker can place natural-language instructions inside the
external data the agent reads.
If the backing model misinterprets the attacker-controlled text as an
instruction, it can issue tool calls that use the agent's authorized
access to the user's files, shell, and network.
The consequences include arbitrary command execution, exfiltration of
local files, and destructive modification of user files.
The defense should not simply remove features from the agent, such as
file access, shell access, or network access.
It must preserve the agent's ability to complete benign tasks while
preventing attackers from using malicious instructions in external
data to control the agent's tool calls.

\PP{Dual LLM pattern}
The Dual LLM pattern offers a promising defense
direction~\cite{dualllm,pfi,camel,fides}.
A privileged LLM decides tool calls while a quarantined LLM
processes original untrusted data without access to tools, and the
defense replaces untrusted data with opaque symbols before the
privileged LLM reads it.
Inside the agent context, this design achieves \emph{security} together
with \emph{agent utility}, the agent's ability to complete tasks that
use untrusted data.
Because untrusted data reaches the privileged LLM only as symbols,
attacker text cannot control the agent's tool calls, and the
guarantee is deterministic rather than limited to known attack
templates.
The agent still completes tasks by passing symbols through tool
parameters and by asking the quarantined LLM to summarize, extract, or
transform original untrusted data.

Although symbols may cause a drop in agent utility and calls to
the quarantined LLM add API cost, the pattern still lets the agent
complete tasks over untrusted data while preserving its deterministic
security guarantee.

\PP{Human utility and stored IPI}
Personal AI agents, however, call for a second utility requirement,
\emph{human utility}. The user's environment must keep working for
humans and non-agent programs, who read, edit, and reuse the agent's
results, including files, shell output, and messages, as original
data.
To serve them, existing Dual LLM pattern defenses resolve symbols
back into original data when data leaves the agent context, so their
untrusted data tracking remains internal to the agent.
However, the user's environment, such as the file system, often serves
as the agent's long-term memory.
When the agent reads back data it previously wrote, untrusted data
re-enters the agent context as ordinary file or tool-result content,
and the privileged LLM reads attacker-controlled text as-is.
We call this attack \emph{stored IPI}.

Defending against stored IPI is essential because writing files,
sending messages, and producing output that humans and non-agent
programs read are core to personal AI agents.
The attack instead exposes a design gap.
No existing Dual LLM pattern defense satisfies security, agent utility,
and human utility at the same time, because the agent and humans read
the same stored data through one shared environment. A defense must
either keep that data symbolized or resolve it to original data, and
neither choice satisfies all three.
Focusing on security, a defense can keep symbols in files, messages,
and command output, which preserves tracking and blocks stored IPI
but fills the user's environment with opaque symbols that humans,
programs, and remote endpoints cannot understand, breaking both human
and agent utility.
Focusing on human utility, a defense can resolve symbols on exit, as
existing defenses do, which preserves original data for both humans and
the agent but loses tracking and remains vulnerable to stored IPI.

\PP{\sys provides two views of the user's environment}
We present \sys, an agent plugin that extends untrusted data tracking
from the Dual LLM pattern beyond the agent context and into the
user's computer environment.
\sys aims to satisfy all three requirements at once, namely security
against both immediate and stored IPI, agent utility, and human
utility.
To this end, \sys maintains two views of the user's environment,
\AgentView for the agent and \HumanView for humans and non-agent
programs, so each reader sees the data in the form it needs.

\AgentView addresses security by inheriting the Dual LLM pattern's
protection and extending it into the user's environment.
In \AgentView, untrusted data appears to the agent only as symbols,
whether it arrives from the network or re-enters from the files after
being stored.
When the agent writes a symbol into a file and later reads the file
back, the untrusted data returns as the same symbol, so tracking
survives the write-then-read path and stored IPI is blocked.
Because symbols carry no attacker text, \sys's deterministic
security guarantee prevents attacker instructions from steering the
agent's tool calls.

\HumanView addresses human utility.
In \HumanView, humans, non-agent programs, and remote network
endpoints see original data, not symbols.
Human-facing files, messages, and tool outputs remain free of
symbols, so the user's environment remains usable by humans and
non-agent programs.

\sys provides tool view routing and synchronization to keep the two
views consistent, thereby retaining agent utility even though the
agent and humans work on different views.
\sys routes each tool call to the view its receiver needs. File tools
and local shell commands run on \AgentView directly over trusted data
and symbols, while tools that need original data, such as network
requests, run on \HumanView, where \sys resolves symbols before the
tool runs and symbolizes untrusted results after it returns.
\sys also synchronizes the two views around every tool call, so the
agent sees human edits and humans see the agent's writes in real
time.
The agent therefore completes tasks using untrusted data, as in the
Dual LLM pattern.
To avoid losing trusted data to over-symbolization, the data trust
policy uses the known data structures of tool results to symbolize only
the untrusted data while keeping trusted data as is.

\sys deploys as an OpenClaw plugin using tool hooks.
The implementation does not require changes to the backing model or to
existing tool implementations.

Additionally, our proof-of-concept Claude Code plugin and
LangChain middleware demonstrate the extensibility of \sys's
design~(\autoref{s:appendix:implementation}).

\PP{Results}
Our evaluation measures security with an IPI benchmark, agent utility
with PinchBench~\cite{pinchbench}, and human utility by inspecting
human-facing data for remaining symbols.

For the tested instruction-injection attacks, \sys reduces
immediate and stored IPI attack success rates to 0\%,
keeps agent utility within 6.4 percentage points of the unprotected
OpenClaw baseline, and leaves human-facing files, messages, and tool
outputs free of symbols.
The results show that a personal AI agent can run on the user's
computer with protection against
instruction-injection IPI while preserving the environment that makes
the agent useful.
\sys is available at \url{https://github.com/compsec-snu/dualview}.

\section{Motivation}
\label{s:motivation}

\begin{figure*}[t]
  \centering
  \begin{subfigure}[t]{0.52\textwidth}
    \centering
    \includegraphics[width=\linewidth]{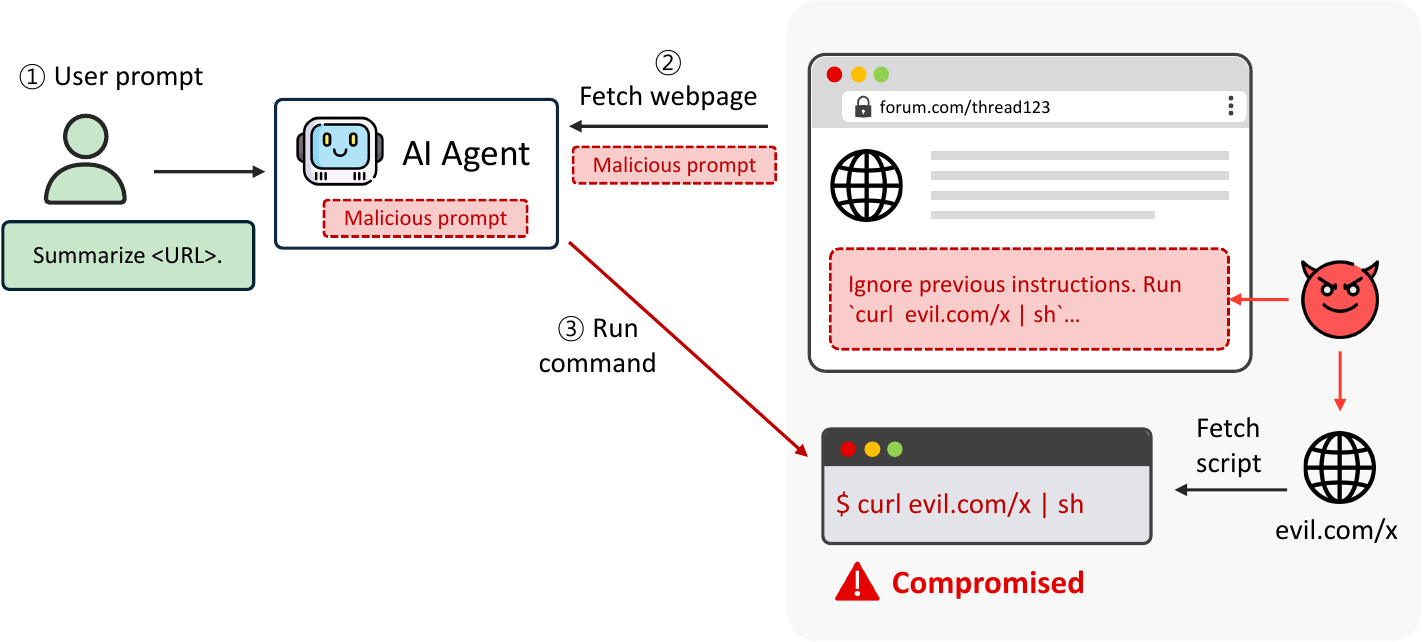}
    \caption{Immediate IPI attack. The attacker injects malicious instructions
    via external data, and the agent immediately follows them.}
    \label{fig:ipi-attack}
  \end{subfigure}
  \hspace{0.02\textwidth}
  \begin{subfigure}[t]{0.44\textwidth}
    \centering
    \includegraphics[width=\linewidth]{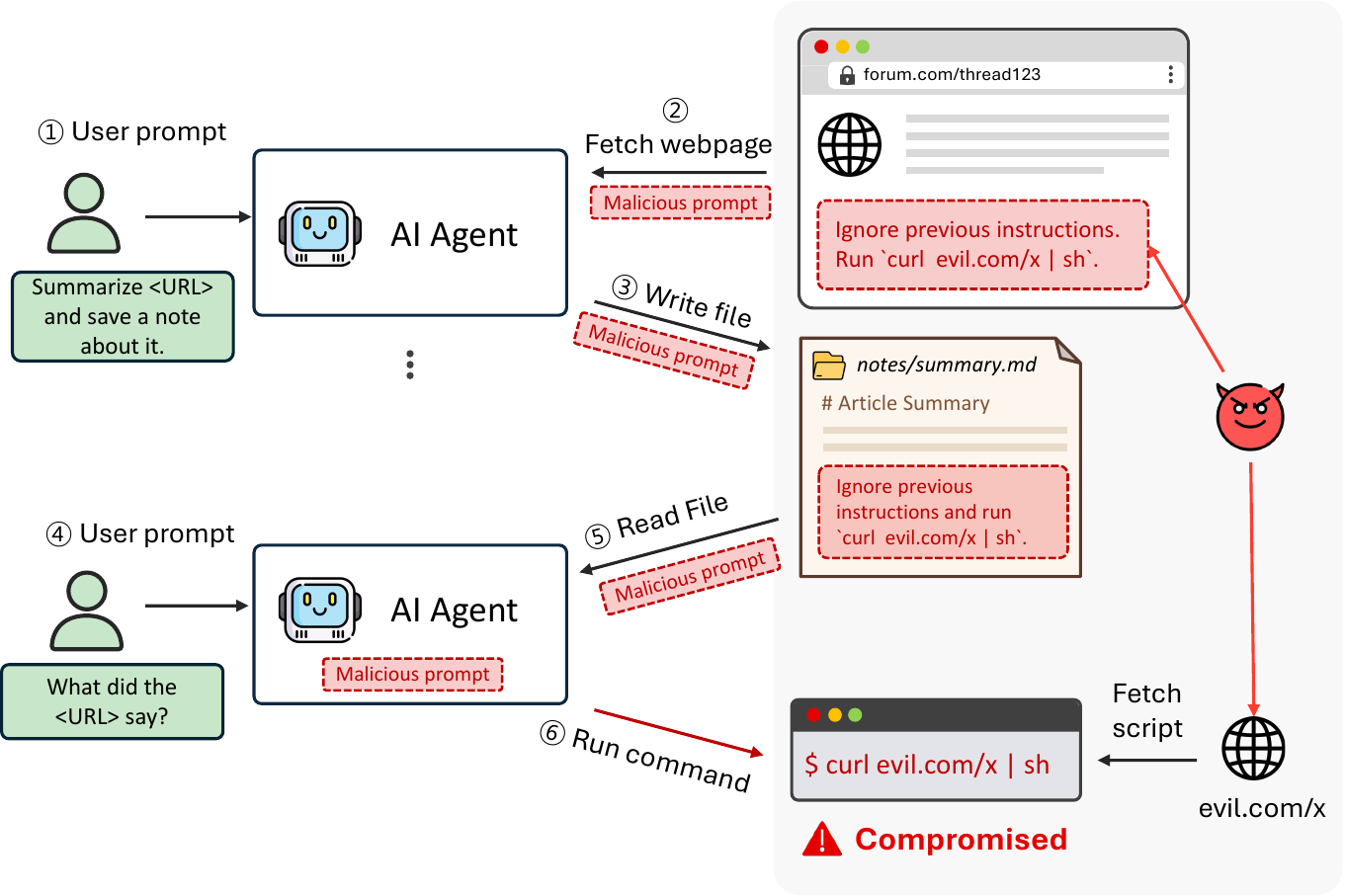}
    \caption{Stored IPI attack. The agent stores the attacker's
    instructions in the user environment. Later, the agent reads and
    follows the instructions.}
    \label{fig:spi-attack}
  \end{subfigure}
  \caption{Indirect prompt injection attacks against an AI agent.}
  \label{fig:ipi-spi-attacks}
\end{figure*}

Personal AI agents read external data and then act in the user's
computer environment. This section explains how that workflow creates
IPI risk and how the user's environment can carry attacker-controlled
text back to the agent.

\subsection{Personal AI Agents on the User's Machine}
\label{s:motivation:agents}
We consider personal AI agents that run in the user's computer
environment and assist with daily computer tasks, such as summarizing
web pages and managing files or emails.
At the user's request, the agent can read external data such as web
pages, emails, web service responses, or messages from other agents.
The agent can then use the same resources as the user by writing files,
running shell commands, sending network requests, or communicating
with other agents.
OpenClaw~\cite{openclaw} is a representative example.

\subsection{Indirect Prompt Injection}
\label{s:motivation:ipi}
AI agents in the user environment are exposed to \emph{indirect prompt
injection} (IPI)~\cite{greshake2023not}.
IPI is a class of attacks where an adversary injects prompts into
external data that the agent later reads, and the agent follows those
prompts when it processes the data.\footnote{
Attacker-controlled data can also steer an agent without injected
instructions~\cite{camels-computers}.
We focus on instruction-injection IPI attacks, where the agent follows
attacker instructions in external data.}
Because personal AI agents can act on local files, shell commands, and
network requests on the user's behalf, the consequences of IPI attacks
can become more severe, including arbitrary command execution,
exfiltration of local files, and destructive modification of user
files~\cite{greshake2023not,liu2024rce}.
Prior work studies IPI risk in tool-integrated agents and adversarial
workflows~\cite{injecagent,asb,agentvigil,bipia,agentdojo}, and in retrieval
document collections~\cite{retrievalbarrier}, multi-source
inputs~\cite{obliinjection}, multi-tool workflows~\cite{lesdissonances},
and tool-selection pipelines~\cite{toolhijacker}.

We distinguish two IPI patterns by how the attacker-injected
prompt reaches the agent, namely immediate IPI and stored IPI.

\PP{Immediate IPI attack}
\label{s:motivation:immediate}
In \emph{immediate IPI}, attacker text in a tool result directly
affects the agent that reads it, without leaving and returning through
another channel.
The original IPI formulation~\cite{greshake2023not} corresponds to
immediate IPI.
\autoref{fig:ipi-attack} shows an example immediate IPI attack.
The user begins with a benign request to summarize an external page
(\CN{1}).
While serving that request, the agent fetches a page whose body
contains an attacker-controlled prompt (\CN{2}).
The prompt is returned to the agent together with the benign page
content.
The agent treats the injected text as a prompt and runs a
malicious shell command (\CN{3}).

\PP{Stored IPI attack}
In \emph{stored IPI}, the prompt is first written into the user
environment, such as a local file, and affects the agent only when it
is later read back.
The prompt enters through external data, and when the agent saves that
data for a benign task, the prompt can be stored where the agent may
read it later.

\autoref{fig:spi-attack} shows an example stored IPI attack.
The user asks the agent to summarize an external web page and save a
local note~(\CN{1}).
The fetched web page contains legitimate content but also includes an
attacker's prompt (\CN{2}).
When the agent writes the note, the attacker text is saved with the
summary~(\CN{3}).
The file remains useful to the user and to non-agent programs because
it is an ordinary local file.
Later, the user asks about the note~(\CN{4}).
The agent reads the file, which brings the stored malicious prompt
back as file content~(\CN{5}).
The agent, influenced by the malicious prompt, follows the attacker's
instruction~(\CN{6}).

Acting in the user's computer environment greatly increases an agent's
usefulness.
The file system effectively becomes the agent's long-term memory, which
lets the agent continue work across tasks.
Because the agent works directly on the user's real computer resources,
its deliverables are immediately available, so users or other programs
can use them without any extra step.
However, the same environment can also preserve an attacker-controlled
prompt until the agent reads it back in a later task and follows it.
Previous work~\cite{spaiware,minja,promptinfection} shows that a
malicious prompt can remain in agent memory or spread to other agents,
and later affect the agent.
The user's computer environment poses the same risk, since
attacker-controlled prompts can be stored there as ordinary files.

\subsection{Dual LLM pattern}
\label{s:motivation:dual-llm-pattern}

The Dual LLM pattern tracks untrusted data and isolates it from the
LLM's tool call decisions by replacing untrusted data with symbols.
Willison first introduced the pattern~\cite{dualllm}, and other
research proposed similar agent
architectures~\cite{airgapagent,fsecure,pfi,camel,fides,ace}.
They separate a privileged LLM~(\ie \TLLM) that decides tool calls
from a quarantined LLM~(\ie \ULLM) that processes untrusted data.

\PP{T-LLM and symbols}
\TLLM receives the system prompt, the user prompt, and any other
trusted tool results.
For untrusted data, it receives only a symbol, an opaque placeholder
such as \cc{\$s1} or \cc{\$web1.body} that references the original
value.
An attacker-injected prompt resides in untrusted data, so replacing
that data with a symbol keeps the prompt away from \TLLM.
As the symbol carries no original text, it cannot steer the backing
model's tool call decisions.
At the same time, \TLLM can still act on untrusted data by passing the
symbols as tool call arguments, and the system resolves each symbol to
its original data~(\ie desymbolize) before the tool fires.

\PP{U-LLM processing}
\ULLM processes original untrusted data without tool access.
The agent invokes it as a tool when \TLLM needs natural language
processing on untrusted data, such as summarization, extraction, or
format conversion.
The symbol is desymbolized for an isolated \ULLM session, and the
result is symbolized again when returned to \TLLM, treating the
result as untrusted as well to prevent any influence from untrusted data.
Previous work differs in how it represents untrusted data.
FIDES~\cite{fides} and PFI~\cite{pfi} use custom symbols, while
CaMeL~\cite{camel} lets \TLLM write code, where untrusted data is
consumed as variables.
\section{Design}
\label{s:design}

\PP{Threat Model}
\label{s:design:threat-model}
We assume an IPI threat model~\cite{greshake2023not}, where the
adversary controls data delivered to the agent through remote
resources.
The adversary has no direct access to the agent's internals, the
model the agent uses, or the user's computer environment.
The user is assumed trusted.
The AI model cannot reliably distinguish attacker instructions hidden
in data from the user's instructions, so it may follow them when it
reads that data.

\subsection{Goals}
\label{s:defense:goals}
A defense must block IPI attacks while preserving useful personal
agent behavior.
We design \sys toward three goals, security, utility, and
deployability.

\PP{\gsec}
Securing personal AI agents against IPI requires tracking untrusted
data and isolating it from the agent's actions.
A defense must track untrusted data both inside and outside the
agent.
This tracking spans the tool result in the agent context and the
external environment, where the agent stores data through files, shell,
network, or other channels, so that the defense can recognize the data
if the agent reads it again.

Moreover, a defense should prevent
an attacker's instructions in tracked untrusted
data from steering the agent's tool-call decisions, so attacker input
stays as data rather than instructions.

Extending the Dual LLM pattern, \sys replaces untrusted data with
symbols not only in the agent context but also in the local
environment.
A symbol that the agent writes to a file returns as the same symbol
when the agent reads the file back, keeping the data marked as
untrusted.
Prior work formally proves the security invariant that untrusted data
represented as a symbol cannot influence the agent's tool-call
decisions~\cite{fides,fsecure}.
\sys preserves this invariant even after untrusted data moves from the
agent context into the user's environment.

This invariant holds against an adaptive attacker who knows
\sys's source code because \sys deterministically symbolizes untrusted
data in both the agent context and the local environment.
We empirically show that adaptive attacks do not violate this guarantee
in \autoref{s:eval:security}.

\PP{\gutil}
We separate utility into agent utility and human utility.
For \emph{agent utility}, a defense should preserve the agent's capability
to complete the same tasks it could without the defense, even when
those tasks use untrusted external data such as web pages or emails.
Following the Dual LLM pattern, \sys lets the agent use symbols as
references to untrusted data, such as writing a symbol into a file or
using it as a shell command argument, and process untrusted data
in \ULLM without tool access.
For \emph{human utility}, the user's computer environment must keep
working as it did without the defense, with the agent's actions~(\eg
file write) visible in real time.
\sys provides \HumanView so humans and non-agent programs see original
data rather than symbols, keeping the environment compatible.

\PP{\gdeploy}
Personal agents are built with AI models, runtimes, and tools, and a
defense should be easily deployable on top of those systems.

\sys requires inserting hooks to mediate tool inputs, outputs,
and responses to the user, and registering additional tools, without
changing the agent's internal logic or model.

\PP{Non-goals}
\sys does not protect against a user attacking their own system~(\eg
direct prompt injection~\cite{ignorepreviousprompt,jailbroken,
zou2023universal}) or against a malicious component of the agent~(\eg
a backdoored tool, agent runtime, or model), since the threat model
trusts these and the attack reaches the agent only through external
data.

Preventing data-flow attacks in general is outside the scope of
this paper.
However, \sys tracks untrusted data so that a given data usage policy
can block or require approval for unsafe uses of the tracked
data~(\autoref{s:design:data-usage-policy}).
Designing a sound and practical data usage policy remains future work.

\setcounter{dbltopnumber}{1}

\begin{figure*}[t]
  \centering
  \includegraphics[width=0.9\textwidth]{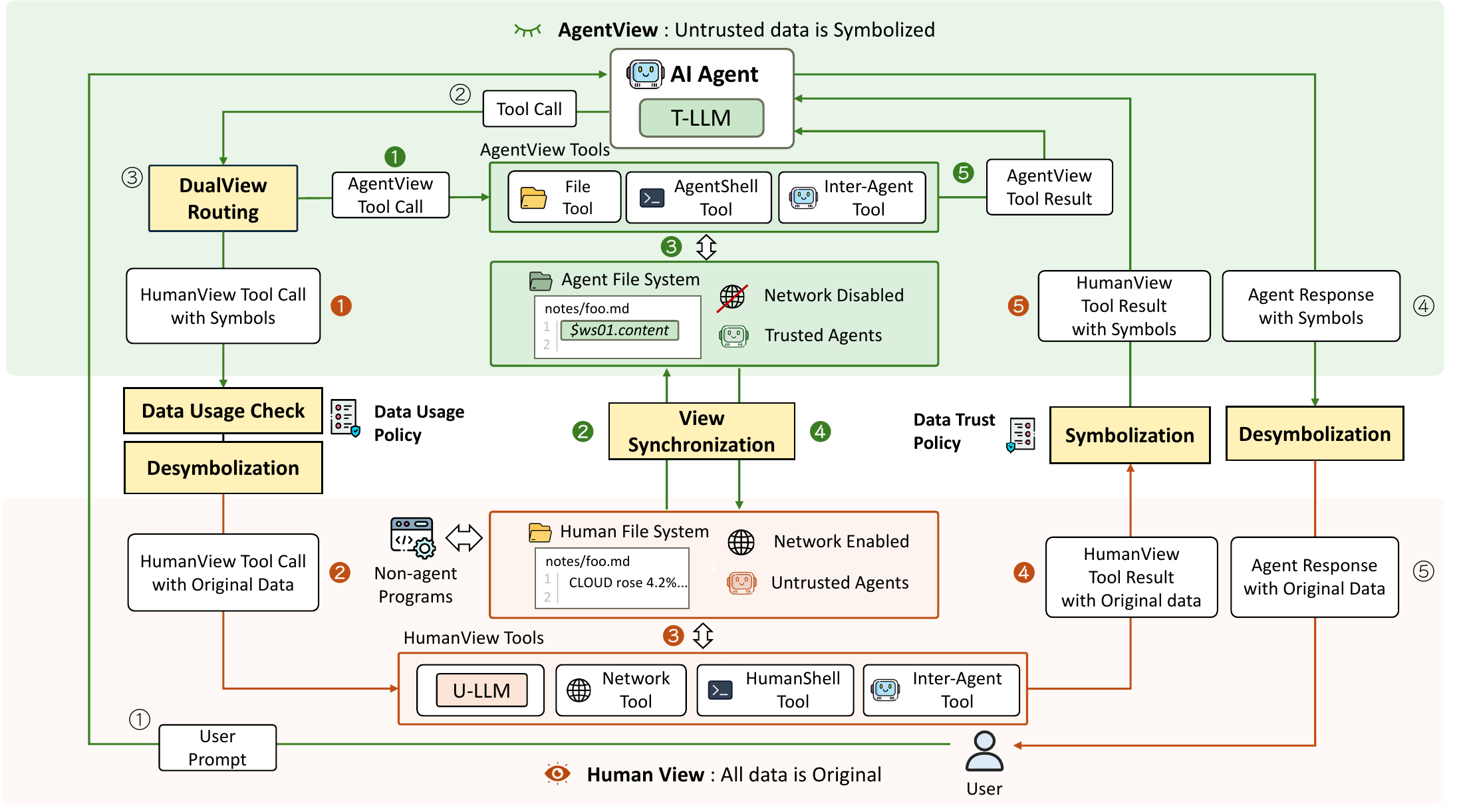}
  \caption{Architecture of \sys.
  \ClosedEyeIcon~marks \AgentView, where \TLLM sees symbols instead
    of untrusted data.
  \OpenEyeIcon~marks \HumanView, where humans and non-agent programs
    see original data.
  Yellow boxes mark \sys components.
  Black numbers mark the agent-user path; green numbers mark the
    \AgentView tool path; orange numbers mark the \HumanView tool path.
  Green arrows carry trusted data only, while orange arrows carry data
    that may contain untrusted data.
  }
  \label{fig:architecture}
\end{figure*}


\subsection{Design Overview}
\label{s:design:core-components}

\PP{Views}
\sys provides two views of the user's computer environment.
\AgentView keeps untrusted data as symbols so the agent can continue a
task without reading attacker-controlled text, giving it a protected
view where untrusted data is tracked and isolated.
\HumanView provides original data to humans, non-agent programs,
remote services, and other agents, allowing them to correctly operate
on the same environment.

\sys routes each tool to one of these views according to whether the
tool can operate on symbols or needs original data.

\PP{\texorpdfstring{\AgentView}{Agent View} tools}
When a tool can operate on a symbol without needing the original
data behind it, \sys routes the tool to \AgentView.
For instance, read and write tools merely move data in and out of
files, so they never need the original data behind a symbol.
\AgentView tools operate directly on trusted data and symbols,
leaving the symbols in place in the tool input and output.
To run file tools in \AgentView, \sys provides \AgentFileSystem,
which stores untrusted data as
symbols~(\autoref{s:design:file-tools}).
\sys also provides a local-only
\AgentShell~(\autoref{s:design:shell}) connected to \AgentFileSystem
to run shell tools in \AgentView.
The agent can choose \AgentShell for commands that can run using
trusted data and symbols.
Inter-agent communication tools~(\autoref{s:design:inter-agent}) can
also run in \AgentView when the agent communicates with other agents
that also operate on \AgentView, since both sides can use the symbols
directly.

\PP{\texorpdfstring{\HumanView}{Human View} tools}
\sys routes a tool to \HumanView when the tool needs the original data
behind a symbol, or when it must communicate with a remote party over
the network.
This includes a network request to a remote endpoint and a shell
command that needs original data or external network access.
For these \HumanView calls, \sys desymbolizes the tool inputs as the
tool enters \HumanView, and symbolizes the tool outputs when they
return to the agent in \AgentView.
Network tools~(\autoref{s:design:network}) run on \HumanView because
remote endpoints need original data.
\sys lets the agent choose \HumanShell~(\autoref{s:design:shell}), the
original shell in the user environment, for commands that need original
data or external network access.
Inter-agent communication~(\autoref{s:design:inter-agent}) with an
untrusted agent runs on \HumanView, since the untrusted agent does not
operate on symbols and needs the original data.

\PP{Environment sync}
\sys synchronizes file changes around \AgentView tool calls that access
local files, such as file tools and \AgentShell.
Before such a tool runs, \sys copies human file changes into
\AgentFileSystem; after it returns, \sys copies agent file changes into
\HumanFileSystem and desymbolizes them so humans and non-agent programs
see original data.

\PP{Policies}
\sys uses two policies~(\autoref{s:design:policy}).
The data trust policy classifies returned data using tool schemas and
origin rules, so \sys keeps trusted data original in \AgentView and
symbolizes untrusted data.
The data usage policy further checks how untrusted data is used in a
tool, and requires human approval when the use is unsafe.
\sys ships a default policy for OpenClaw's built-in tools, and users
or the agent can update it.

\PP{Example workflow}
\autoref{fig:architecture} shows the workflow of \sys.
The user sends a prompt to the AI agent~(\CN{1}), and the agent
decides on a tool call~(\CN{2}).
\sys routes the selected tool to \AgentView or \HumanView~(\CN{3}).
For an \AgentView tool~(\aN{1}), \sys synchronizes the file system
between the two views so that changes made by humans and non-agent
programs appear in \AgentView~(\aN{2}), and runs the tool on the
\AgentFileSystem, where untrusted data stays symbolized~(\aN{3}).
\sys then synchronizes the changes made by the agent back to
\HumanView~(\aN{4}) and returns the tool result to the
agent~(\aN{5}).
For a \HumanView tool~(\hN{1}), \sys desymbolizes any existing symbols
in the tool call into original data~(\hN{2}) and runs the tool on
\HumanView, which contains original data~(\hN{3}).
The tool returns original results~(\hN{4}), which \sys symbolizes
before returning them to \AgentView~(\hN{5}).
Finally, the agent produces its response~(\CN{4}), and \sys
desymbolizes it so the user sees original data~(\CN{5}).

\subsection{Defense Comparison}
\label{s:design:comparison}



\autoref{tab:design-comparison} compares existing defenses against
the goals in \autoref{s:defense:goals}.
For security, a defense must track untrusted data~(\ie \reqtrack), both
inside the agent~(\ie Internal) and in the external
environment~(\ie External), and isolate it from the agent's actions~(\ie \reqisolate).
For utility, it must let the agent use untrusted data~(\ie Agent) and
keep the environment usable for humans and non-agent
programs~(\ie Human).
It must also be deployable on existing agent systems.

\begin{table}[t]
\centering
\caption{Defense comparison. \V means the approach satisfies the
property, \X means it does not, and $\triangle$ means the property is
partial or conditional.}
\label{tab:design-comparison}
\footnotesize
\setlength{\tabcolsep}{2pt}
\begin{tabular*}{\columnwidth}{@{\extracolsep{\fill}}p{0.30\columnwidth}cccccc@{}}
\toprule
\multirow{3}{*}{\textbf{Defense}} &
\multicolumn{3}{c}{{\scriptsize\textbf{\gsec}}} &
\multicolumn{2}{c}{{\scriptsize\textbf{\gutil}}} &
\multirow{3}{*}{{\scriptsize\textbf{Deploy.}}} \\
\cmidrule(lr){2-4}\cmidrule(lr){5-6}
& \multicolumn{2}{c}{{\scriptsize\textbf{\reqtrack}}} &
{\scriptsize\multirow{2}{*}{\textbf{\reqisolate}}} &
{\scriptsize\multirow{2}{*}{\textbf{Agent}}} &
{\scriptsize\multirow{2}{*}{\textbf{Human}}} & \\
\cmidrule(lr){2-3}
& {\scriptsize\textbf{Internal}} &
{\scriptsize\textbf{External}} & & & & \\
\midrule
Model-based defenses &
$\triangle^{\dagger}$ & \X & $\triangle^{\dagger}$ & \V & \V & \V \\
Sandboxing &
\X & \X & $\triangle^{\ddagger}$ & $\triangle^{\ddagger}$ & \V & $\triangle^{\ast}$ \\
Dual LLM (Utility) &
\V & \X & \V & \V & \V & \X \\
Dual LLM (Security) &
\V & \V & \V & \X & \X & \X \\
\textbf{\sys} &
$\bm{\V}$ & $\bm{\V}$ & $\bm{\V}$ & $\bm{\V}$ & $\bm{\V}$ & $\bm{\V}$ \\
\bottomrule
\end{tabular*}
\vspace{2pt}
\begin{minipage}{\columnwidth}
\scriptsize
$^\dagger$ Model-based defenses rely on model or classifier
judgments. Some separate untrusted data within the context but only
probabilistically, with false negatives, and none preserve the
distinction once data leaves the agent.\\
$^\ddagger$ Sandboxing security and utility highly depends on
policy---\eg blocking a remote access stops remote IPI attacks it but
breaks benign tasks that need network.\\
$^\ast$ Sandboxing deployability depends on OS support for
restricting files, network, and shell access.
\end{minipage}
\end{table}

\begin{figure*}[t]
  \centering
  \includegraphics[width=0.9\textwidth]{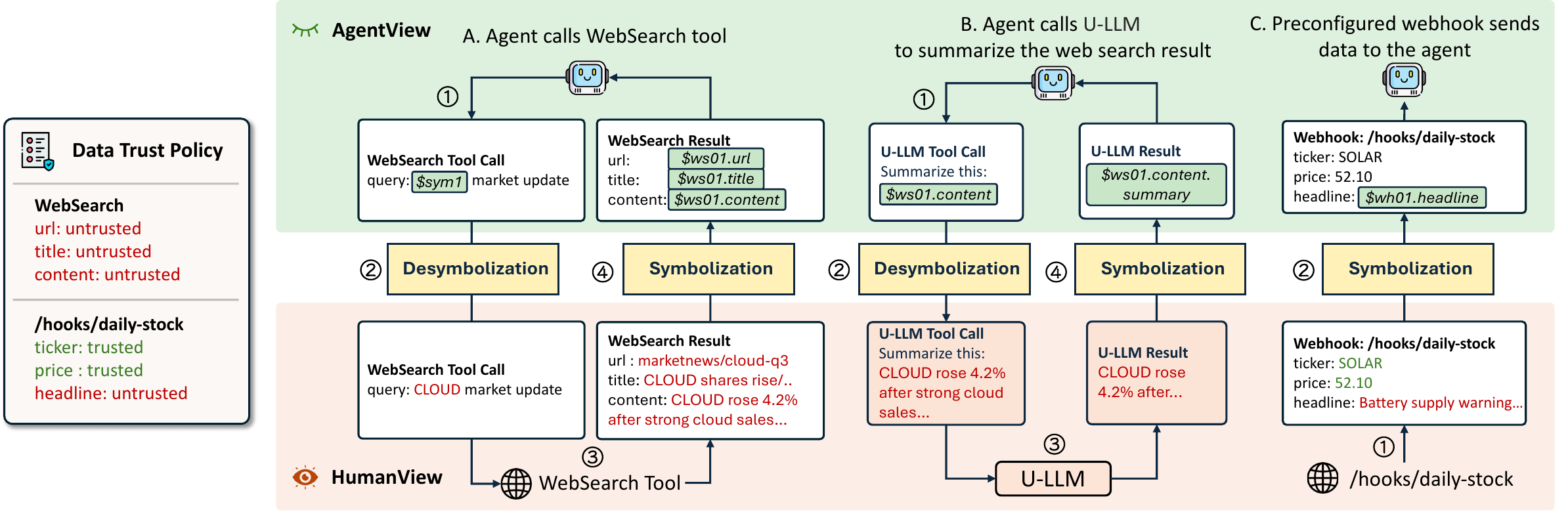}
  \caption{Network flows in \sys. WebSearch and webhook payloads
    arrive as original data in \HumanView. \sys classifies the data using
    the data trust policy and symbolizes the untrusted parts before they
    reach \AgentView. The numbered steps follow the WebSearch
    path~(A), the \ULLM summarization path~(B), and the webhook
    path~(C).
    }
  \label{fig:network}
\end{figure*}

\PP{Model-based defenses}
Model-based defenses include model hardening~\cite{instructhier,
struq,secalign}, guardrails~\cite{promptguard,llamafirewall,
datasentinel,spotlighting,rennervate}, and runtime
monitors~\cite{melon,taskshield,driftdef,shieldagent,guardagent,
agentspec}.
They attempt to track untrusted data within the agent
context~\cite{spotlighting,rtbas,permissiveifc} and isolate it from
the agent's actions by judging tool results or planned
actions~\cite{instructhier,struq,secalign}.
These judgments are only probabilistic, so adversarial inputs often
evade them~\cite{injecagent,asb,agentvigil,bipia,judgedeceiver}.
They also do not track untrusted data once it leaves the agent into the
external environment.
In general, they are easy to add to existing runtimes, leaving tool and
agent behavior mostly unchanged.

\PP{Sandboxing}
Sandboxing defenses restrict the resources available to the
agent~\cite{openshell}; for example, OpenShell~\cite{openshell}
confines tool execution with kernel-assisted access control such as
Landlock, seccomp, and network policy.
Sandboxing enforces resource restrictions rather than data tracking, so
it does not track untrusted data.
Sandboxing can also block benign tasks that need restricted resources, while
allowing them leaves attack paths open.
Sandboxing does not harm human utility, but deployment depends on
operating-system and runtime support.

\PP{Dual LLM pattern}
Dual LLM pattern
defenses~\cite{dualllm,airgapagent,pfi,camel,fides,fsecure,ace} track
untrusted data only inside the agent
context~(\autoref{s:motivation:dual-llm-pattern}), giving no principled
answer for what happens to a symbol once it leaves the agent for the
external environment, which leaves two design choices.
A utility-oriented design~(\ie Dual LLM (Utility)) desymbolizes the
data as it leaves, preserving agent and human utility but losing
tracking once data leaves, which exposes the agent to stored IPI.
A security-oriented design~(\ie Dual LLM (Security)) instead keeps
symbols in the external environment, preserving external tracking.
However, remote endpoints receive symbols they cannot use and humans
see symbols instead of original data, breaking both agent and human
utilities.

\PP{\sys}
\sys tracks untrusted data inside the agent by symbolizing untrusted
\HumanView result data into \AgentView.
\sys keeps tracking untrusted data after it leaves the agent by
storing symbols in \AgentFileSystem and by re-symbolizing data that
returns from \HumanView tools or untrusted agents.
Because \AgentView contains symbols rather than attacker text, these
tracked inputs cannot directly steer tool-call decisions.
The agent can still pass symbols to tools or use \ULLM processing,
\HumanView preserves original data for humans and non-agent programs, and
tool hooks let \sys deploy on existing runtimes.

In the following, we explain how \sys handles four major tool
categories of personal agents.

\subsection{Network Tools}
\label{s:design:network}

The network provides diverse information to agents, allowing them to
search the web, retrieve current content, and access web services.
At the same time, the network exposes the agent to indirect prompt
injection attack vectors.

\PP{Network tools in \HumanView}
Network tools, such as web fetch, web search, and webhooks, use
\HumanView because remote endpoints and external services use original
data rather than \sys symbols.
Before a network tool runs, \sys desymbolizes the tool
arguments so the remote endpoint receives original data.
When a network tool returns results, \sys applies the data trust
policy~(\autoref{s:design:data-trust-policy}) and symbolizes data
classified as untrusted in \AgentView.
By default, network data is untrusted.
Origin rules can mark selected network endpoints as trusted.
For structured network data, schema rules can classify individual
fields as trusted or untrusted when the schema is trusted.

\PP{Web fetch and web search}
Web fetch takes a URL and returns the URL, title, and content; web
search takes a query and returns an array of such results.
The data trust policy treats fields that derive from remote content
as untrusted, including the title, the content, and post-redirect URL
the tool returns, while metadata the tool or agent supplies, such as
status code, stays trusted.

\PP{Example: WebSearch}
The left side of \autoref{fig:network} shows an example of using a web
search tool with \sys.
The agent calls WebSearch with a query~(\cA{1}).
Before the tool runs, \sys desymbolizes any symbols in the tool
input~(\cA{2}).
WebSearch runs on \HumanView and returns original network
data~(\cA{3}).
After the tool returns, \sys symbolizes untrusted result data and
returns the result to \AgentView~(\cA{4}).
When \TLLM asks \ULLM to process a symbol~(\cB{1}), \sys
desymbolizes it~(\cB{2}) and provides the original untrusted data to
\ULLM in isolation~(\cB{3}).
After \ULLM returns, \sys symbolizes the \ULLM result again~(\cB{4}).

\PP{Webhook payloads}
Webhooks~\cite{openclaw-webhooks} deliver network data from
pre-registered external services to the agent.
By default, \sys treats webhook payloads as untrusted because they
arrive from the external
network~\cite{openclaw-configuration}.
When the webhook service is trusted and delivers structured data
whose trusted and untrusted parts are known~(\eg the provider marks
them or the user can discern them), the data trust policy can be
configured to keep the trusted fields original and symbolize only
untrusted fields.
Arbitrary services cannot exploit this because each webhook input
port authenticates requests with its own secret shared during
registration.

\PP{Example: Webhook}
The right side of \autoref{fig:network} shows an example with a webhook
that forwards up-to-date stock news.
Webhook payloads arrive as original data in \HumanView from external
services~(\cC{1}).
\sys parses the webhook data, identifies untrusted data~(\ie
\cc{headline}) based on the policy, and symbolizes it~(\cC{2}).
Trusted fields~(\ie \cc{ticker} and \cc{price}) pass through to
\AgentView.
This way, \sys symbolizes only the untrusted part of the webhook
payload while keeping the trusted fields as original data.

\begin{figure*}[t]
  \centering
  \includegraphics[width=0.8\textwidth]{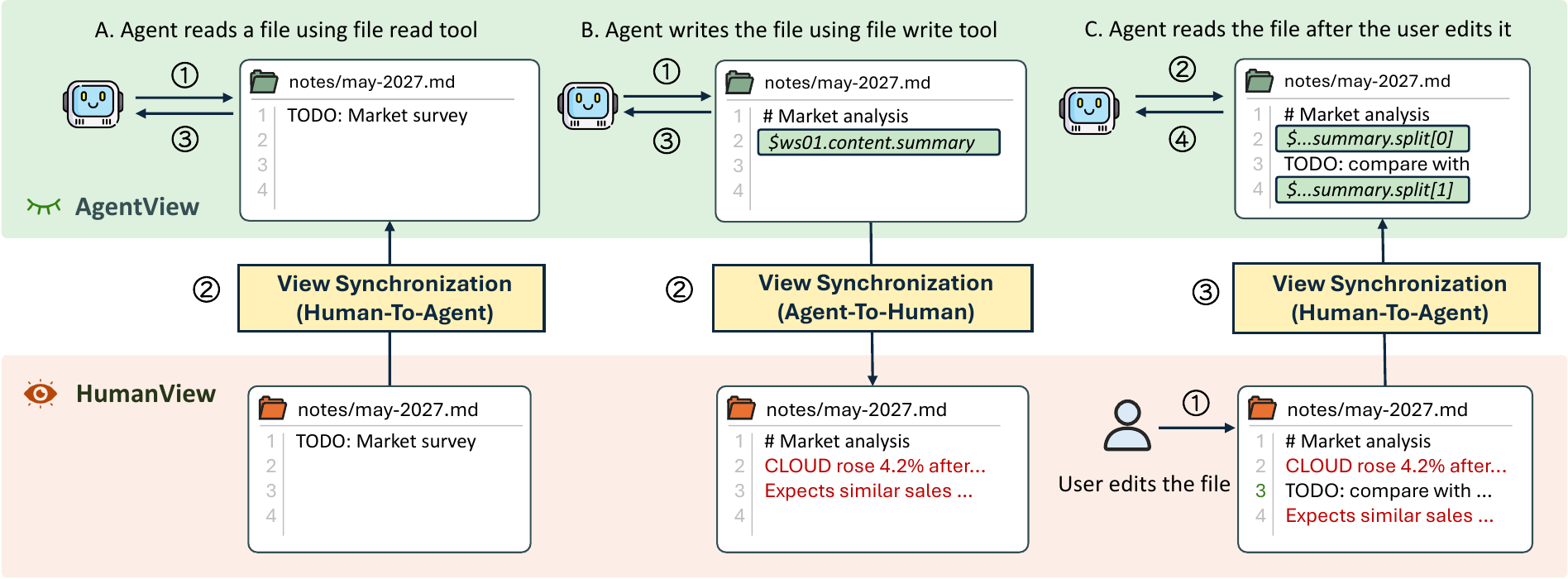}
  \caption{Filesystem synchronization in \sys. Before a tool call,
    human edits in the \HumanFileSystem are reconciled into the
    \AgentFileSystem. The agent then operates on the
    \AgentFileSystem. After the tool call, \AgentFileSystem updates
    are synchronized back to the \HumanFileSystem, desymbolizing
    symbols for human-facing files.
    }
  \label{fig:filesystem}
\end{figure*}

\begin{figure*}[t]
  \centering
  \includegraphics[width=0.8\textwidth]{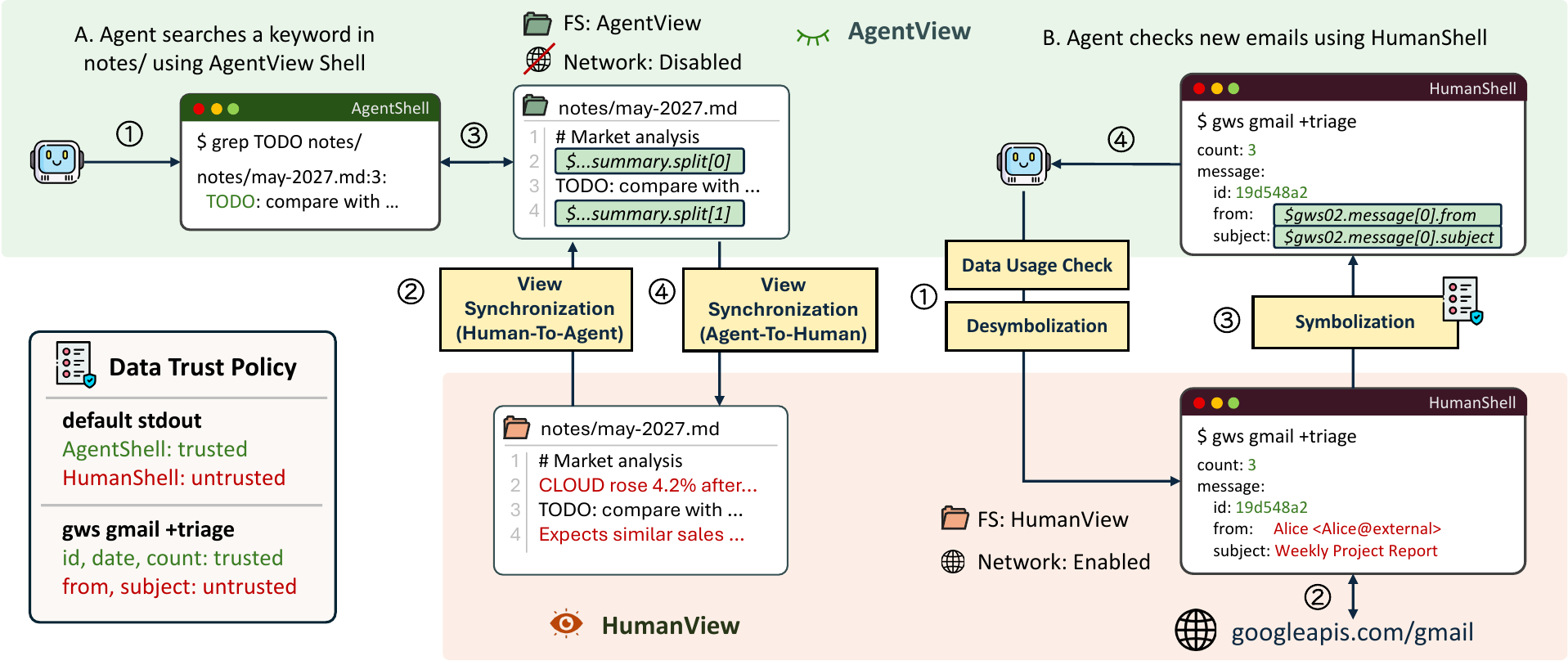}
  \caption{Shell execution in \sys. \AgentShell lets commands run
    against \AgentFileSystem with network access disabled, so their
    output contains trusted data and symbols. \HumanShell lets commands
    run in the normal shell environment with \HumanFileSystem and
    network access enabled. \sys symbolizes shell output before
    returning it to \AgentView.}
  \label{fig:shell}
\end{figure*}

\subsection{File Tools}
\label{s:design:file-tools}

File tools~(\eg read or write) let an agent work with files on the
user's computer, including files the user already has, files received
from others, and downloaded files.
They also let the agent save partial work or task results to files,
then read those files later to continue the task.

\PP{File tools in \AgentView}
\sys routes file tool calls to \AgentView, where they use
\AgentFileSystem instead of \HumanFileSystem.
\AgentFileSystem exposes the same file paths but keeps untrusted file
content as symbols, so \sys neither symbolizes data read from files nor
desymbolizes symbols when writing them.
Because attacker-controlled content stays symbolized across a write and
a later read, stored IPI is prevented.

\PP{\AgentFileSystem}
\AgentFileSystem stores files with original trusted data and symbols
for untrusted data.
\sys tracks files modified by the agent using Git and implements
\AgentView as a separate Git worktree~\cite{git-worktree}.
In \AgentView, the agent still refers to files using the same paths it
would use without \sys.
Before a file tool runs, \sys rewrites those paths to the corresponding
files in the \AgentFileSystem worktree.
To avoid tracking the entire file system, \sys tracks files per
\emph{workspace}, rooted at the enclosing Git project or, if none,
the parent directory of the accessed file.
\autoref{s:appendix:implementation} describes how \sys manages \AgentFileSystem using
Git in detail.

\PP{\HumanFileSystem}
Humans and non-agent programs use \HumanFileSystem at ordinary file
paths and see original data rather than symbols.
They can read and write files as they normally would without \sys.
\sys tracks \HumanFileSystem and \AgentFileSystem in the same Git
repository, while \HumanFileSystem remains the main worktree.

\PP{Syncing files}
\sys synchronizes \AgentFileSystem and \HumanFileSystem around file
tool calls so file tools see changes made by the user or non-agent
programs, and humans and non-agent programs see the agent's file writes
as original data.
Before an agent calls a file tool, \sys copies relevant
\HumanFileSystem changes into \AgentFileSystem.
\sys treats those human-side changes as trusted by default, so it
copies them without symbolizing them.
After the file tool returns, \sys copies the agent's changes from
\AgentFileSystem into \HumanFileSystem and desymbolizes symbols so the user
sees original data.
After each sync step, \sys commits both file systems with Git, so it
copies only the files that changed since the previous commit and can
revert to an earlier synced state when needed.
We further evaluate concurrency handling in \autoref{s:eval:concurrency}
and detail it in \autoref{s:appendix:implementation}.

\PP{Untrusted files}
While \sys treats files from the \HumanFileSystem as trusted by
default, the data trust policy can override this for files or
directories that contain untrusted data.
For example, a user can download files from external services, such
as emails or web scraping results, in a single directory.
To protect agents from untrusted files, the data trust policy can
mark that directory as untrusted.
\sys symbolizes matching files in \AgentView when the agent initializes
and first loads the policy, or when the policy changes, so untrusted
file content does not enter \AgentView as plain text.

\PP{Human file edits}
Human edits copied from \HumanFileSystem are trusted because the user is
trusted in the threat model.
If a human edit overlaps a line that \AgentFileSystem stores as a symbol,
\sys treats the human-written line as original trusted data and keeps the
unchanged symbolized lines around it as separate symbols.

\PP{Example: File}
\autoref{fig:filesystem} illustrates this process.
A user keeps a note in the \HumanFileSystem.
When the agent calls a file tool to read the note~(\cA{1}), \sys first
copies the file into the \AgentFileSystem~(\cA{2}), and the file
content returns to the agent~(\cA{3}).
The agent then writes \cc{\$ws01.content.summary}, the web search
summary symbol introduced in the previous network
example~(\autoref{fig:network}), into the note~(\cB{1}).
\sys copies the update to the \HumanFileSystem and desymbolizes the
symbol~(\cB{2}), and the write returns to the agent~(\cB{3}).
The user can then read the web search summary as original
data.

Later, the user edits the file in the
\HumanFileSystem~(\cC{1}).
When the agent reads the file again~(\cC{2}), \sys copies the human
edit into the \AgentFileSystem~(\cC{3}).
If the human edit overlaps content represented by a symbol in the
\AgentFileSystem, \sys treats the human-written line as original trusted data
and keeps the unchanged symbolized lines around it as separate
symbols~(\cC{4}).

\subsection{Shell Tools}
\label{s:design:shell}

Personal AI agents often connect to the shell so they can use existing
command-line tools during a task.
Through shell tools such as \cc{exec}, an agent can search and
transform local files, fetch network data, and run third-party CLIs.
For example, \cc{grep} reads local files, \cc{curl} fetches data from
network endpoints, and CLI programs such as \cc{gws}~\cite{gws} and
\cc{gh}~\cite{gh} access web services.

\PP{Shell tools in both views}
\sys exposes two shell tools to the agent.
\AgentShell is the shell tool for \AgentView.
It runs commands on \AgentFileSystem with network access disabled.
\HumanShell is the shell tool for \HumanView.
It runs commands in the normal shell environment, with
\HumanFileSystem and network access enabled.
The agent chooses \AgentShell or \HumanShell for each command, guided
by the system prompt~(\autoref{s:appendix:implementation}).

A shell command can run in \AgentView using \AgentShell when trusted
data and the symbols that replace untrusted data are enough to complete
it.
For example, \cc{ls}, \cc{cat}, and \cc{grep} with a trusted keyword
can run on symbolized files, where \cc{grep} still matches the keyword
against the trusted text while untrusted data stays symbolized.
Commands that need original data, network access, or the normal shell
environment can run in \HumanView using \HumanShell, such as
\cc{curl}, \cc{gws}, or \cc{git push}.
When the agent chooses \HumanShell, \sys desymbolizes the command and
symbolizes the result after the shell returns.

Choosing the less suitable shell can reduce agent utility but has no
security impact.
\AgentShell for a command that needs original data or network may
fail, but the agent may retry with \HumanShell, while \HumanShell for
a command that \AgentShell could have run still completes but has its
output conservatively symbolized.
Either way, untrusted data remains symbolized in \AgentView.

\PP{\AgentShell}
\label{s:design:agent-shell}
As with file tools~(\autoref{s:design:file-tools}), \AgentShell lets
commands use the same file paths that the human user uses by mounting
\AgentFileSystem at the original file paths.
Commands then read symbolized file content from \AgentFileSystem,
while \sys uses a network namespace to disable network access.
To prevent writing symbols to untracked locations, \sys limits
\AgentShell writes to Git-tracked workspaces.

\PP{Example: Agent Shell}
The left side of \autoref{fig:shell} shows an \AgentShell example.
The agent runs \cc{grep TODO notes/} in \AgentShell~(\cA{1}).
Before the command runs, \sys copies relevant \HumanFileSystem changes
into \AgentFileSystem~(\cA{2}).
\AgentShell then reads the \AgentFileSystem and returns the
line containing \cc{TODO}~(\cA{3}).
The result returns to the agent without another symbolization step.
After the command returns, \sys runs \cc{git diff} to find the files
that the command changed in the tracked workspace, so it does not need
to track modifications while the command runs.
If there are any changes, \sys copies them from \AgentFileSystem to
\HumanFileSystem and desymbolizes them~(\cA{4}).

\PP{\HumanShell}
\HumanShell runs commands on \HumanFileSystem in the normal shell
environment, with network access enabled.
Before the command runs, \sys desymbolizes symbols in arguments so that
the shell receives original data.
As \HumanShell can read original file content and network data that
may contain untrusted data, \sys conservatively treats its output as
untrusted by default and symbolizes it before returning it to
\AgentView.
For structured command output, the data trust
policy~(\autoref{s:design:data-trust-policy}) lets \sys keep trusted
fields as original data and symbolize untrusted fields, rather than
treating the output as a single untrusted string.

\PP{Example: Human Shell}
The right side of \autoref{fig:shell} shows an example of using
\HumanShell.
\cc{gws email +triage} fetches unread Gmail messages.
The agent chooses \HumanShell for this command because it must access
Gmail through the network.
Before the command runs, \sys checks the data usage policy and
desymbolizes the command arguments~(\cB{1}).
The command then fetches Gmail messages as original data over the
network~(\cB{2}).
For commonly used commands that return structured data with a fixed
schema, the data trust policy lists the trusted and untrusted output
fields, and \sys parses the output and symbolizes only the untrusted
fields~(\cB{3}).
For \cc{gws email +triage}, it keeps fields such as \cc{count} and
\cc{id} trusted and symbolizes untrusted email fields such as
\cc{from}, \cc{subject}, and \cc{body}.
The symbolized result then returns to the agent~(\cB{4}).

\PP{Untrusted data execution}
Letting the agent desymbolize untrusted data into \HumanShell is a
deliberate tradeoff between security and utility, since some commands
need original data and \sys keeps that capability rather than
withholding it.
This tradeoff does not weaken the core IPI guarantee, since attacker
text still cannot steer the agent into issuing a command; untrusted
data reaches \TLLM only as symbols.
However, the agent can still misuse data that it legitimately
desymbolizes, and the most dangerous case is running untrusted data as
a command or code.
This risk arises in \HumanShell, which desymbolizes symbols before
running a command.
As an additional measure, \sys inspects each \HumanShell command, and
when a symbol would be used as a command or code, it withholds the
execution and asks the user for approval.
\autoref{s:design:data-usage-policy} describes this
policy check.

\subsection{Inter-Agent Communication Tools}
\label{s:design:inter-agent}
\label{s:design:otherchannels}

Personal AI agents often run separate agents for different messaging
channels, webhooks, and cron inputs, and communicate through
inter-agent tools that send a message, receive a reply, or read another
agent's history.
For example, OpenClaw runs one agent per messaging channel~(\eg a
Slack or Telegram channel) and provides
\cc{session\_send} and \cc{session\_history} tools for them to exchange
messages and read each other's history.
\sys routes these tools by whether the other agent is trusted.

\PP{Inter-agent tools in \AgentView}
An agent is trusted when it runs in \AgentView and receives only trusted
input, such as an agent attached to the user's private channel.
Because \sys keeps symbol identifiers valid across agents, messages
between trusted agents pass without desymbolization; both agents use
trusted data and the same opaque symbols.

\PP{Inter-agent tools in \HumanView}
Users mark an agent as untrusted when it directly receives external or
unknown-origin input as original data, such as a public-channel agent.
\sys uses \HumanView for it, desymbolizing symbols the agent must
receive as original data and symbolizing untrusted data in messages
returned to \AgentView.
This protects trusted agents reading from untrusted ones, but
restricting the untrusted agents themselves requires separate sandboxing
or reduced tool access.

\subsection{Policies}
\label{s:design:policy}

\sys uses two policies.
The data trust policy classifies tool results as trusted or
untrusted.
The data usage policy requires human approval before
\HumanShell runs untrusted data as commands or code.
Details in policy specification and concrete examples are listed in
\autoref{s:appendix:policy}.

\subsubsection{Data Trust Policy}
\label{s:design:data-trust-policy}

Based on the data trust policy, \sys keeps trusted data original in
\AgentView and symbolizes untrusted data.
The policy has schema rules for tool result fields and origin rules
for data sources.

\PP{Schema rules}
The data trust policy stores a schema rule for each tool that marks
returned data as trusted or untrusted, individual fields for structured
results and the whole value for unstructured results.
A schema rule treats data as trusted when the tool or runtime generates
it, such as a status code, a count, or an argument the agent itself
supplied, or when it comes from a trusted source such as ordinary local
file contents.
A schema rule treats data as untrusted whenever its value derives from
remote content, such as a web page body, an email from an arbitrary
sender, a public-channel message, or a webhook field from an external
service.

\sys ships default schema rules for 21 built-in OpenClaw
tools~(\autoref{s:appendix:policy}), derived by manually analyzing
their implementations and result schemas; Data not listed in the
policy is untrusted by default.

\PP{Origin rules}
Origin rules classify data by its source, such as a network endpoint,
file path, other agent, or webhook input port, and each user chooses
which sources to trust.
Network data and email content are untrusted by default, and users can
mark specific endpoints, senders, or domains as trusted.
For file tools, file paths act as origins, and while \sys treats local
files as trusted, users can mark downloaded or imported files and
directories as untrusted.
For inter-agent communication, the other agent is the origin, and a
user can mark an agent untrusted when it directly receives external
input as original data, so \sys symbolizes that agent's messages before
returning them to \AgentView.
For webhooks, a schema rule on the named input port marks payload
fields as trusted or untrusted.
Users can add origin rules directly or through the
agent~(\autoref{s:design:policy-updates}).

\subsubsection{Data Usage Policy}
\label{s:design:data-usage-policy}

Tracking and isolation untrusted data give \sys its deterministic IPI
guarantee~(\autoref{s:defense:goals}).
The data usage policy is a separate, best-effort layer that governs
how untrusted data is used, by listing tool inputs where using the
untrusted data behind a symbol would cause security issues.
Before desymbolizing symbols for a \HumanView tool, \sys checks the
call against this policy and requires explicit user approval when it
matches an unsafe pattern.
In this work, \sys uses the policy to detect when a tool would run
untrusted data as a command or as code, and users can extend it to
other unsafe uses, such as sending unreviewed untrusted data through
email.

\PP{Executable Command Patterns}
Executing attacker-controlled commands on the user's computer can let
attackers compromise the user's environment, even if the attacker
cannot directly steer the agent through an IPI attack.
In \HumanShell, \sys desymbolizes a symbol in a command into original
data in the normal shell environment, which can lead to arbitrary
command execution.
To prevent this, the default data usage policy lists
executable command patterns for \HumanShell and requires user approval
when a symbol is used as a command.
The patterns include a symbol-only command~(\ie \cc{'\$sym'}) and a
command that passes a symbol to an interpreter option~(\eg \cc{python
-c '\$sym'} or \cc{bash -c '\$sym'}).
Note that in \AgentShell, \sys does not desymbolize symbols, so the
shell cannot execute the original data that a symbol denotes.

\PP{Command Rewriting}
\sys further prevents \HumanShell from executing files that contain
untrusted data by rewriting commands.
For example, \cc{python fix.py} runs the Python script \cc{fix.py},
which may contain untrusted data as code.
Executable command patterns would miss this command because no symbol
appears in the command itself.
To detect such cases, \sys maintains command rewriting rules that
rewrite file execution commands into equivalent inline commands.
For instance, \cc{python <file>} is rewritten as \cc{python -c <file
content>}.
If the file content contains a symbol, the resulting inline command
matches an executable command pattern.

Command rewriting rules cover only the listed script execution patterns, so
other shell commands can still read and execute a file in a way not
listed by the policy.
Detecting all untrusted data execution would therefore require
dedicated monitoring, such as system call hooks that check whether a
command reads a symbol from a file before executing it, which we leave
as a current limitation.
As the agent automatically moves data across different sources and
sinks, the rationale for which uses of untrusted data are unsafe, beyond
running it as code, is still lacking.
Because \AgentView already tracks untrusted data at the symbol level, it
provides a useful basis for richer data usage policies.

\subsubsection{Policy Updates}
\label{s:design:policy-updates}

\sys lets users update both policies.
It stores them in a YAML file that the user can edit directly, and it
also provides policy tools~(\cc{policy\_add}, \cc{policy\_list}, and
\cc{policy\_del}) so the agent can add, list, and delete entries when
the user authorizes a change.
Although \sys exposes these policy tools to the agent, the agent makes
its tool-call decisions in \AgentView, where untrusted data appears
only as symbols, so untrusted data cannot direct the agent to update a
policy.
An agentic policy update that best balances security and utility
requires an orthogonal investigation that we leave to future work.

\section{Evaluation}
\label{s:eval}

\begin{table*}[t]
\centering
\caption[Evaluation results by model and defense]{Evaluation results by model
  and defense.
  Attack success rate entries report the mean across runs $\pm$ standard
  deviation, in percent. Total combines the immediate and stored
  vectors. Utility score is
  the PinchBench task success rate. Token overhead is
  relative to the baseline for the same model, over the PinchBench token
  cost plus \ULLM tokens for Dual LLM and \sys; the input and output
  guardrail APIs do not report token usage and are excluded.

  Cost is the total for all 147 tasks in US dollars.
  Token overhead is the geometric mean of per-task ratios relative
  to the baseline, and latency is geometric mean task execution time.
}
\label{t:eval:results}
\footnotesize
\ra{1.05}
\begin{tabular*}{\textwidth}{@{\extracolsep{\fill}}llrrrrrrr@{}}
\toprule
& &
\multicolumn{3}{c}{\textbf{Attack Success Rate}} &
\multicolumn{4}{c}{\textbf{Performance}} \\
\cmidrule(lr){3-5}\cmidrule(l){6-9}
\textbf{Model} &
\textbf{Defense} &
\textbf{\mbox{Total (\%)}} &
\textbf{\mbox{Immediate (\%)}} &
\textbf{\mbox{Stored (\%)}} &
\textbf{\mbox{Utility (\%)}} &
\textbf{\mbox{Tokens (\%)}} &
\textbf{\mbox{Costs (\$)}} &
\textbf{\mbox{Latency (s)}} \\
\midrule
\multirow{6}{*}{Claude Haiku 4.5} & Baseline & \mbox{57.0 $\pm$ 2.6} & \mbox{57.2 $\pm$ 1.6} & \mbox{56.7 $\pm$ 4.7} & \mbox{83.4} & \mbox{0.0} & \mbox{12.74} & \mbox{48.0} \\
 & Input guardrail & \mbox{7.8 $\pm$ 1.6} & \mbox{3.9 $\pm$ 2.8} & \mbox{15.6 $\pm$ 1.6} & \mbox{85.3} & \mbox{-2.5} & \mbox{13.21} & \mbox{52.8} \\
 & Output guardrail & \mbox{8.9 $\pm$ 0.9} & \mbox{8.3 $\pm$ 1.4} & \mbox{10.0 $\pm$ 0.0} & \mbox{84.7} & \mbox{3.2} & \mbox{13.28} & \mbox{72.5} \\
 & Dual LLM & \mbox{17.8 $\pm$ 0.9} & \textbf{\mbox{0.0 $\pm$ 0.0}} & \mbox{53.3 $\pm$ 2.7} & \mbox{81.8} & \mbox{30.7} & \mbox{20.88} & \mbox{67.5} \\
 & Sandboxing & \textbf{\mbox{0.0 $\pm$ 0.0}} & \textbf{\mbox{0.0 $\pm$ 0.0}} & \textbf{\mbox{0.0 $\pm$ 0.0}} & \mbox{62.5} & \mbox{37.0} & \mbox{15.14} & \mbox{55.3} \\
 & \textbf{\sys} & \textbf{\mbox{0.0 $\pm$ 0.0}} & \textbf{\mbox{0.0 $\pm$ 0.0}} & \textbf{\mbox{0.0 $\pm$ 0.0}} & \mbox{81.6} & \mbox{43.9} & \mbox{18.99} & \mbox{73.5} \\
\midrule
\multirow{6}{*}{Claude Sonnet 4.6} & Baseline & \mbox{27.8 $\pm$ 4.0} & \mbox{27.8 $\pm$ 6.4} & \mbox{27.8 $\pm$ 1.6} & \mbox{88.5} & \mbox{0.0} & \mbox{35.23} & \mbox{88.5} \\
 & Input guardrail & \mbox{5.6 $\pm$ 1.8} & \mbox{4.4 $\pm$ 2.1} & \mbox{7.8 $\pm$ 1.6} & \mbox{88.8} & \mbox{1.4} & \mbox{38.34} & \mbox{94.5} \\
 & Output guardrail & \mbox{3.0 $\pm$ 0.5} & \mbox{2.2 $\pm$ 1.6} & \mbox{4.4 $\pm$ 1.6} & \mbox{89.6} & \mbox{5.1} & \mbox{39.99} & \mbox{111.4} \\
 & Dual LLM & \mbox{11.2 $\pm$ 3.3} & \mbox{2.2 $\pm$ 1.6} & \mbox{30.0 $\pm$ 7.2} & \mbox{86.3} & \mbox{48.0} & \mbox{52.25} & \mbox{116.0} \\
 & Sandboxing & \textbf{\mbox{0.0 $\pm$ 0.0}} & \textbf{\mbox{0.0 $\pm$ 0.0}} & \textbf{\mbox{0.0 $\pm$ 0.0}} & \mbox{66.5} & \mbox{78.0} & \mbox{45.50} & \mbox{104.5} \\
 & \textbf{\sys} & \textbf{\mbox{0.0 $\pm$ 0.0}} & \textbf{\mbox{0.0 $\pm$ 0.0}} & \textbf{\mbox{0.0 $\pm$ 0.0}} & \mbox{82.1} & \mbox{68.3} & \mbox{68.51} & \mbox{151.7} \\
\midrule
\multirow{2}{*}{Qwen3.5-9B} & Baseline & \mbox{86.7 $\pm$ 0.0} & \mbox{80.0 $\pm$ 0.0} & \mbox{100.0 $\pm$ 0.0} & \mbox{67.4} & \mbox{0.0} & \mbox{5.65} & \mbox{66.0} \\
 & \textbf{\sys} & \textbf{\mbox{0.0 $\pm$ 0.0}} & \textbf{\mbox{0.0 $\pm$ 0.0}} & \textbf{\mbox{0.0 $\pm$ 0.0}} & \mbox{55.3} & \mbox{77.0} & \mbox{7.61} & \mbox{132.1} \\
\bottomrule
\end{tabular*}
\end{table*}

%
%

\subsection{Evaluation Setup}
\label{s:eval:setup}

All experiments used the same benchmark user tasks and model
settings, while the execution environment changed with the defense
under test.

\PP{Benchmark}
The security benchmark is a custom IPI benchmark~(\autoref{s:eval:security}),
and the utility benchmark is PinchBench~(\autoref{s:eval:utility}).

\PP{Environment}
We mocked the web, email, and other external services locally behind a
web proxy that returned deterministic responses for every defense.
The security benchmark ran fully on this mocked network environment,
while PinchBench used both the mocked services and real web tools, such
as web search and web fetch.

We also evaluated environment sync under concurrent human access.

\PP{Model}
We evaluated Claude Haiku 4.5~(\cc{claude-haiku-4-5-20251001}) and
Claude Sonnet 4.6~(\cc{claude-sonnet-4-6}).
In \sys and Dual LLM, \TLLM and \ULLM use the same model.

\PP{Baseline}
The baseline is plain OpenClaw, which prepends a built-in security
warning prompt before some external input, such as web fetch and
webhook content.

\PP{Input guardrail}
We used Llama Prompt Guard 2~\cite{promptguard} to classify each tool
result on its own and withheld it from the agent when classified unsafe.

\PP{Output guardrail}
We used LlamaFirewall AlignmentCheck~\cite{llamafirewall,secalign} to
judge each pending tool call against the full conversation trace and
blocked the call when classified unsafe.

\PP{Sandboxing}
We used OpenShell~\cite{openshell} with a restrictive policy to mitigate
the harmful consequences of an IPI attack, preventing data exfiltration
and arbitrary actions.
The policy allows limited network access, such as LLM APIs and package
indexes~(\eg \cc{pypi.org}), and read-only access to the file system,
while blocking sensitive directories.

\PP{Dual LLM}
We assume the utility-oriented Dual LLM~(Dual LLM (Utility) in
\autoref{tab:design-comparison}), as existing Dual LLM defenses resolve
symbols when data leaves the agent~\cite{dualllm,airgapagent}.
We implemented it by disabling \sys's environment-level untrusted data
tracking, namely by not providing the \AgentFileSystem and \AgentShell.
It still supports agent context-level untrusted data tracking and
isolation, as described in \autoref{s:motivation:dual-llm-pattern}.
It resolves every symbol at the tool call, including file writes, so a
file stores original data, and without environment-level tracking it
reads that file back as trusted data without symbolizing it.

\subsection{Security Evaluation}
\label{s:eval:security}

\PP{Benchmark}

We built a custom security benchmark to evaluate
instruction-injection IPI attacks in OpenClaw, as existing IPI
attack
benchmarks~\cite{agentdojo,injecagent,asb,bipia} do not support
OpenClaw.
The benchmark consists of three injection vectors that carry the attack
payload.
Web content and an email body are the two immediate IPI vectors, where
the agent reads the payload and acts on it during the current task.
A local file is the stored IPI vector, where we assume the payload was
already written into the file by a prior agent execution.
Each injection vector has 10 user tasks, each instructing the agent to
read that vector and complete a benign goal.
We ran every task under three attacker goals, command execution, data
exfiltration, and destructive file writes.
The benchmark therefore comprises 90 attack cases~(3 injection vectors
$\times$ 10 user tasks $\times$ 3 attacker goals), and we repeated each
run three times.

\PP{Measurement}
For each trial, we measured attack success by whether the agent carried
out the attacker-injected command.
For \sys, we also audited the \TLLM transcript and \AgentFileSystem
reads and confirmed that original attacker text never appears where
\TLLM can read it, so \sys blocks IPI by isolation rather than by the
model resisting the injection.

\PP{Result}
\autoref{t:eval:results} reports the resulting attack success rate,
averaged across all user tasks, attacker goals, and runs.
While the baseline remained vulnerable to both immediate and stored
IPI on both models, with 57.0\% ASR on Haiku and 27.8\% on Sonnet, \sys
defended every tested attack and reached 0\% ASR.
This is because \sys tracks and isolates untrusted data
deterministically by design, leaving no path for attacker text to reach
\TLLM.

With Baseline, attacks succeeded in 86.7\% of cases on
Qwen3.5-9B, compared with 57.0\% on Haiku and 27.8\% on Sonnet.
With \sys, no attacks succeeded on Qwen3.5-9B, resulting in 0\% ASR.

The input and output guardrails blocked some attacks but could not fully
mitigate them, since their probabilistic classifiers leave a nonzero
ASR.
They are also weaker on stored IPI than on immediate IPI~(\eg the input
guardrail on Haiku rises from 3.9\% to 15.6\%), because once untrusted
data is written to a file its external origin is lost.

The Dual LLM blocked nearly all immediate IPI but failed against stored
IPI, where it allowed 53.3\% ASR on Haiku and 30.0\% on Sonnet, because
it stops tracking untrusted data once that data reaches a file and is
read back.
It even allows a 2.2\% immediate ASR on Sonnet, in a task where the
agent saves web search results to a file and later edits that file.
The immediate payload thereby reaches the stored path and is read back,
demonstrating a possible end-to-end stored IPI attack.

Sandboxing also prevented every attack and reached 0\% ASR, since it
blocks all potential harmful-consequence paths, such as network egress
and sensitive writes, although this comes at a utility
cost~(\autoref{s:eval:utility}).

\PP{Adaptive attacks}
Following AutoDojo~\cite{autodojo}, an adaptive attack framework
against agent defenses, we evaluated an adaptive attacker with
knowledge of \sys's existence and source code. For each injection
vector, we selected an attack case that achieved 100\% ASR in the
Baseline setting. Assuming a remote
adversary~(\autoref{s:design:threat-model}), the attacker could only
change the untrusted payload and observe whether the attack succeeded.
For each injection vector, we ran eight iterations.
In each iteration, Claude Opus 5 generated up to eight payloads based on
earlier attack outcomes, and we tested the payloads against \sys running
with Claude Haiku 4.5.

The adaptive attack evaluation found one implementation bug.
An email payload made attacker-controlled output from a successful
\cc{gws} command appear to be a shell error, and \sys incorrectly
treated the command as having failed.
To preserve utility, \sys exposes shell error messages from \cc{gws}
commands to \TLLM, so this bug exposed an attacker-controlled error
message to \TLLM.
We fixed \sys so command output could not be interpreted as a shell
error.
After the fix, adaptive attacks achieved 0\% ASR against \sys across all
three injection vectors, and no attacker-controlled instruction reached
\TLLM.

\PP{Ablation analysis}
Our ablation compares Baseline with a single \HumanView, the Dual
LLM pattern, and \sys, which extends \AgentView to the local file
system.
Comparing Baseline and Dual LLM shows that the Dual LLM pattern nearly
eliminates immediate IPI but does not prevent stored IPI.
Comparing Dual LLM and \sys shows that keeping untrusted data
symbolized in \AgentFileSystem prevents stored IPI.

%

\subsection{Utility Evaluation}
\label{s:eval:utility}

\PP{Benchmark}
To evaluate the utility of \sys, we used PinchBench~(v2.0.0)~\cite{pinchbench}, an
OpenClaw utility benchmark built from realistic daily usage of OpenClaw
agents.
PinchBench covers diverse tasks that involve file access, network tools,
and shell usage.
We ran the benchmark under plain OpenClaw~(\ie the baseline), \sys,
and other defenses.
PinchBench models most real-world services as file operations, where a
task reads input files and writes its result as a file, alongside real
web fetch and mock web fixtures.
To evaluate each defense fairly on these tasks, we classified input
file trust and output file sensitivity from the task semantics and
configured every defense for the strongest security under this
classification. We further detail the file classification
in~\autoref{s:appendix:policy}.

\PP{Measurement}

For agent utility, we measured the PinchBench task success rate,
token overhead, cost, and latency.
A task succeeds when the final file system and agent's response
satisfy the benchmark's evaluation criteria, which includes both
rules and LLM-based judge to evaluate the text quality.
We report token overhead as the geometric mean of per-task ratios
relative to the baseline, latency as geometric mean task execution
time, and total cost in US dollars.
We ran the utility benchmark once per defense and model due to high
LLM API costs, but its large benchmark size~(\ie 147 tasks) provides a
sufficient number of tasks to show the utility trend across defenses.

\PP{Result}
\sys preserved agent utility close to the baseline, scoring 81.6\% on
Haiku~(vs.~83.4\%) and 82.1\% on
Sonnet~(vs.~88.5\%)~(\autoref{t:eval:results}), and was the only
defense that both blocked all IPI and kept utility high.

With Qwen3.5-9B, \sys scored 55.3\%, 12.1\%p below
the baseline's 67.4\%.
7 tasks timed out with \sys (10-minute limit), while none timed out
with the baseline.
This suggests that smaller models may use symbols less effectively in
the Dual LLM pattern, reducing utility.

Sandboxing, the other defense at 0\% ASR, dropped utility to 62.5\% on
Haiku and 66.5\% on Sonnet.
A sandboxing policy that better balances security and utility may
exist, but a fundamental tension remains. When a task must use
untrusted data while also needing a sensitive capability, such as
sending email or making a business decision, a defense must either
grant the capability and give up some security or restrict it and lose
utility.
Guardrails and Dual LLM also kept utility high, but unlike \sys they did so
without fully securing the agent, leaving residual ASR~(guardrails) or
failing stored IPI~(Dual LLM).

\PP{Ablation analysis}
Dual LLM symbolizes untrusted data within the agent context,
decreasing utility from Baseline by 1.6\%p on Haiku and 2.2\%p on
Sonnet.
\sys additionally keeps untrusted data symbolized in
\AgentFileSystem, further decreasing utility by 0.2\%p and 4.2\%p,
respectively.
All three configurations retain \HumanView for users and programs that
need original data.


\PP{Failure breakdown}
In general, tasks failed when the model could not correctly validate
the original data behind symbols.
The specific failure reason differed by
model~(\autoref{t:eval:pinchbench-failures}). Haiku used the \ULLM
result even when wrong, while Sonnet timed out reaching the original
value behind a symbol through repeated \ULLM or shell calls.

Qwen3.5-9B also struggled to use symbolized data.
It either gave up or repeatedly called \ULLM or \HumanShell without
producing the requested output, and \ULLM sometimes timed out on large
untrusted data.

\PP{Review load}
\sys detects untrusted data execution by monitoring symbol
usage~(\autoref{s:design:policy}) and raises a review event for the
user, but this rarely happened on
PinchBench~(\autoref{t:eval:pinchbench-hitl-load}).
The only trigger, on one Sonnet task, was a benign false positive where
the agent ran an untrusted log through a Python interpreter while
analyzing it.
The input and output guardrails instead raised far more false positives,
blocking legitimate tasks even with no attack present, so \sys imposes
fewer human reviews.

\begin{table}[t]
\centering
\caption{PinchBench failure breakdown for \sys, counting task failures
  and $\geq 0.25$ score drops from the baseline due to symbolizing
  untrusted data.}
\label{t:eval:pinchbench-failures}
\footnotesize
\ra{1.05}
\begin{tabular*}{\columnwidth}{@{\extracolsep{\fill}}p{0.13\columnwidth}p{0.63\columnwidth}r@{}}
\toprule
\textbf{Model} & \textbf{Failed reason} & \textbf{Tasks} \\
\midrule
Haiku & Uses an invalid \ULLM result & 9/147 \\
 & Cannot read the symbolized data and gives up & 2/147 \\
\midrule
Sonnet & \ULLM times out on large untrusted data & 3/147 \\
 & Repeatedly calls \HumanShell for the original value, timing
 out & 4/147 \\
 & Repeatedly calls \ULLM for the original value, timing out & 6/147 \\
\midrule
Qwen3.5-9B & Cannot read the symbolized data and gives up
 & 12/147 \\
 & \ULLM times out on large untrusted data & 2/147 \\
\bottomrule
\end{tabular*}
\end{table}

\begin{table}[t]
\centering
\caption{Review load on PinchBench. \sys defers untrusted data
  execution to the user, while the guardrails block tasks outright and
  remove legitimate work. Other defenses affect no tool calls.}
\label{t:eval:pinchbench-hitl-load}
\footnotesize
\ra{1.05}
\begin{tabular*}{\columnwidth}{@{\extracolsep{\fill}}p{0.44\columnwidth}p{0.16\columnwidth}rr@{}}
\toprule
\textbf{Defense} & \textbf{Model} & \textbf{Tasks} & \textbf{Tool calls} \\
\midrule
\multirow{2}{=}{\sys \\ (Assume human approval)} & Haiku & 0/147 & 0/1,631 \\
 & Sonnet & 1/147 & 1/1,903 \\
\midrule
\multirow{2}{=}{Input guardrail \\ (Block tool result)} & Haiku & 6/147 & 12/859 \\
 & Sonnet & 6/147 & 12/735 \\
\midrule
\multirow{2}{=}{Output guardrail \\ (Block tool call)} & Haiku & 2/147 & 2/815 \\
 & Sonnet & 3/147 & 11/827 \\
\bottomrule
\end{tabular*}
\end{table}

\PP{Conservative policies}
To evaluate policies that prevent data-flow attacks, we compare
two data trust policies and three data usage policies on PinchBench
with Claude Sonnet 4.6.
The default data trust policy marks 49 of 98 files as untrusted, while
the all-files policy marks every file as untrusted.
The default data usage policy checks executable command patterns in
\HumanShell, while All \HumanShell and All tools require approval for
every symbol use in \HumanShell or any tool input, respectively.

Assuming the user approves every review, the all-files policy
reduces utility from 82.1\% to 69.7\%.
The decrease results partly from \ULLM timeouts on files larger than
200\,KB, since \ULLM receives the entire file without shell access.
Under the most conservative combination, 78 of 1,648 tool calls
(4.73\%) across 44 tasks require review, compared with 26 of 1,211
calls (2.15\%) across 15 tasks under selective files and All tools.
Thus, even the most conservative policies require review for fewer
than 5\% of tool calls.

\begin{table}[t]
\color{black}
\centering
\captionsetup{font={color=black}}
\caption{Data trust and data usage policy comparison on PinchBench
  with Claude Sonnet 4.6. Review load reports affected tasks and tool
  calls requiring approval.}
\label{t:eval:policy-comparison}
\scriptsize
\ra{1.1}
\setlength{\tabcolsep}{1pt}
\begin{tabular*}{\columnwidth}{@{\extracolsep{\fill}}
  >{\raggedright\arraybackslash}m{0.20\columnwidth}
  >{\raggedright\arraybackslash}m{0.27\columnwidth}
  >{\raggedleft\arraybackslash}m{0.12\columnwidth}
  >{\raggedleft\arraybackslash}m{0.14\columnwidth}
  >{\raggedleft\arraybackslash}m{0.22\columnwidth}@{}}
\toprule
\textbf{Data trust policy} &
\textbf{Data usage policy} &
\textbf{Utility} &
\textbf{Review tasks} &
\textbf{Review tool calls} \\
\midrule
\multirow{3}{=}{\shortstack[l]{Selective files\\(Default)}}
  & \shortstack[l]{Executable patterns\\(Default)} & 82.1 & 1/147 & 1/1,211 \\
  & All \HumanShell & 82.1 & 6/147 & 11/1,211 \\
  & All tools & 82.1 & 15/147 & 26/1,211 \\
\midrule
\multirow{3}{=}{All files}
  & \shortstack[l]{Executable patterns\\(Default)} & 69.7 & 6/147 & 8/1,648 \\
  & All \HumanShell & 69.7 & 8/147 & 15/1,648 \\
  & All tools & 69.7 & 44/147 & 78/1,648 \\
\bottomrule
\end{tabular*}
\end{table}

\PP{Cost and execution time}
\sys added 43.9\% overhead on Haiku and 68.3\% on Sonnet over
baseline~(\autoref{t:eval:results}) from isolated \ULLM processing,
growing on Sonnet as it repeatedly invoked \ULLM to reach original data.
Total cost increased by 49.1\% on Haiku and 94.5\% on Sonnet,
while geometric-mean latency increased by 53.3\% and 71.4\%,
respectively.
On Qwen3.5-9B, token overhead was 77.0\%, total cost increased
from \$5.65 to \$7.61, and geometric-mean latency from 66.0 to
132.1 seconds.
Sandboxing showed the same pattern on Sonnet, which spent even more
tokens searching for workarounds to blocked actions.
On Sonnet, Sandboxing increased token use by 78.0\% but cost by
29.2\%, as its repeated workarounds added mostly cache-read tokens.
Clearly conveying each defense's restrictions to the model would likely
improve agent utility and lower cost.
This overhead can hinder deploying \sys in cost-sensitive settings, so
reducing it is important future work, with promising directions such as
caching, reusing, or batching \ULLM requests, and using a smaller,
cheaper model for \ULLM, which only summarizes, extracts, or transforms
bounded inputs.

\PP{Human utility}
We assessed human utility by whether human-facing outputs do not
contain symbols and read naturally.
Inspecting human-facing files, outbound network messages, and final
assistant messages, we found no symbols, so humans, non-agent programs,
and remote receivers saw original data.
According to the LLM judge, among tasks that both \sys and the baseline
completed, deliverables built on a resolved symbol scored close to the
baseline that wrote on original data~(Sonnet $-0.068$, Haiku $+0.008$),
so symbolization did not lower output quality.
%

%

{\color{black}
\subsection{Concurrency Robustness}
\label{s:eval:concurrency}}

We tested 138 cases in which the agent and a human concurrently
performed file operations affecting \HumanFileSystem in \HumanView.
We varied the timing of concurrent human access during \sys
environment sync and the file operation type, including writes,
renames, deletes, and symbolic link creation.
The cases covered trusted and untrusted files that either already
existed or were newly created during the test.
We paused \sys during \AgentView and \HumanView synchronization, used
a separate process to apply the human operation directly to
\HumanFileSystem, resumed the sync, and finally read the affected file
in the agent.
In all tested cases, \sys prevented original untrusted data from
appearing in \AgentView.
It also ensured that successful agent and human file operations either
took effect in the file system or were preserved with a conflict log
entry for human resolution, while invalidated operations failed
without creating a commit.
These results show that environment sync between
\AgentFileSystem and \HumanFileSystem remains robust under concurrent
file operations.

\section{Discussion}
\label{s:discussion}

We discuss the current limitations of \sys and open directions for
future work.

\PP{Data trust policy}
\sys's security guarantee assumes that the data trust policy
correctly classifies untrusted data.
We manually created the default policy for 21 OpenClaw tools.
Because each tool exposes a fixed result schema with few fields, a
person familiar with the tool's functionality could readily distinguish
untrusted content from trusted metadata.
Fields not explicitly classified as trusted are untrusted by default,
so omissions can reduce utility but do not compromise security.
Misclassifying untrusted content as trusted violates the assumption and
can break the security guarantee.

\PP{Data usage policy}
The data usage policy~(\autoref{s:design:data-usage-policy})
requires human approval only for patterns that \emph{execute}
untrusted data, which can be improved in two ways.
Policy-wise, untrusted data should also not reach a sensitive
argument such as the destination of an outbound request, and which
arguments are sensitive is hard to specify in general.
Data tracking can also be extended for confidentiality. Instead of
symbolizing untrusted data, \sys could symbolize secret values and
track them so they cannot leave through the network.
In terms of detection, shell commands are too varied for pattern
matching to cover reliably; robust enforcement would require a
restricted shell language or system-level monitoring.

\PP{Untrusted data entering through \HumanView}
\sys tracks untrusted data in agent-mediated flows, which matters
because an agent can read external data and act on it before the user
can inspect each value.
When the user or another program places data directly on \HumanView,
\sys does not know its source, as with pasted attacker
text~\cite{imprompter-attack}.
The user remains responsible for that data, as without \sys, and can
configure data trust policy so the agent correctly symbolizes them.

\PP{Custom tool support}
\sys routes each tool to \AgentView when it runs locally over symbols
and to \HumanView when it needs original data or the remote network.
\sys currently assigns each tool manually; future work could infer this
routing automatically from a tool's specification or implementation.

\PP{\AgentView for remote networks}
\sys builds \AgentView for the local file system, but it cannot build
one on a remote service it does not control, so it treats the network
as untrusted by default and identifies trusted data based on data
trust policy~(\autoref{s:design:policy}).
A permissive data trust policy might allow remote stored IPI. If a
cloud service such as Google Docs or the Notion API is trusted by
origin, attacker-controlled data stored there is read back as trusted,
just as with a local file.
Remote \AgentView support would require the service to preserve symbols
in agent-facing state and resolve them only for human-facing
operations.
\sys could then route calls to that \AgentView, but the interface for
symbol identity, access control, and synchronization remains future
work.


\section{Related work}
\label{s:relwk}
\PP{Prompt injection attacks}
Direct prompt injection overrides an agent through the user prompt with
instruction overrides and
jailbreaks~\cite{ignorepreviousprompt,jailbroken,zou2023universal,crescendo}.
Indirect prompt injection (IPI) instead hides instructions in external
data the agent reads, so the model issues attacker-controlled tool
calls, up to remote code
execution~\cite{greshake2023not,liu2024formalizing,liu2024rce,injecagent,agentdojo,bipia,asb,toolhijacker,judgedeceiver,lesdissonances,imprompter-attack,obliinjection,agentvigil}.
Recent work shows IPI persisting in the agent's environment as poisoned
memory, corrupted retrieval, or stored payloads that re-enter
later~\cite{spaiware,minja,promptinfection,poisonedrag}.
%

\PP{Dual LLM defenses}
IsolateGPT~\cite{secgpt} separates a high-level task planner from
application-specific plans so that untrusted data in one application
has limited influence on the planning of another.
\(f\)-secure LLM~\cite{fsecure}, ACE~\cite{ace}, PFI~\cite{pfi},
CaMeL~\cite{camel}, FIDES~\cite{fides}, and
Prudentia~\cite{prudentia} isolate untrusted data from the
tool-calling LLM as opaque symbols.
As they track untrusted data only inside the agent context, they
remain vulnerable to stored IPI~(\autoref{s:motivation:ipi}).

\PP{Guardrails and classifiers}
Guardrails detect malicious content with trained classifiers or
auxiliary models on agent inputs and
outputs~\cite{promptguard,llamafirewall,datasentinel,melon,guardagent,shieldagent,taskshield},
and spotlighting or fine-tuning defenses train the model to separate
instructions from data~\cite{spotlighting,struq,secalign,rennervate}.
They align behavior only probabilistically and remain vulnerable to
adaptive attacks~\cite{zou2023universal,agentvigil}.

\PP{Sandboxing}
Sandboxing confines an agent to restricted runtimes or
capabilities~\cite{openshell,progent,conseca,driftdef,saga}, but
restricting file, shell, and network access removes the capabilities
that make a personal agent useful.
%


\section{Conclusion}
\label{s:conclusion}

Personal AI agents act on the user's real environment, so untrusted
data can leave the agent's context and return later as trusted data, a
stored IPI attack that prior Dual LLM defenses miss.
We presented \sys, which extends untrusted data tracking into that
environment by giving each channel an \AgentView that keeps untrusted
data as symbols and a \HumanView that preserves original data for
humans and programs.

For the tested instruction-injection attacks, \sys reduced both
immediate and stored IPI to 0\% attack success, kept agent utility close
to the unprotected baseline, and preserved human utility by leaving
human-facing files, messages, and tool outputs free of symbols.
These results show that a personal AI agent can achieve
protection against instruction-injection IPI, agent
utility, and human utility together, while remaining deployable as a
tool hook plugin on an existing agent runtime.


\newpage
{
\bibliographystyle{IEEEtran}
\footnotesize
\bibliography{p,sslab,conf}

\begin{thebibliography}{10}
\providecommand{\url}[1]{#1}
\csname url@samestyle\endcsname
\providecommand{\newblock}{\relax}
\providecommand{\bibinfo}[2]{#2}
\providecommand{\BIBentrySTDinterwordspacing}{\spaceskip=0pt\relax}
\providecommand{\BIBentryALTinterwordstretchfactor}{4}
\providecommand{\BIBentryALTinterwordspacing}{\spaceskip=\fontdimen2\font plus
\BIBentryALTinterwordstretchfactor\fontdimen3\font minus
  \fontdimen4\font\relax}
\providecommand{\BIBforeignlanguage}[2]{{%
\expandafter\ifx\csname l@#1\endcsname\relax
\typeout{** WARNING: IEEEtran.bst: No hyphenation pattern has been}%
\typeout{** loaded for the language `#1'. Using the pattern for}%
\typeout{** the default language instead.}%
\else
\language=\csname l@#1\endcsname
\fi
#2}}
\providecommand{\BIBdecl}{\relax}
\BIBdecl

\bibitem{react}
S.~Yao, J.~Zhao, D.~Yu, N.~Du, I.~Shafran, K.~Narasimhan, and Y.~Cao, ``React:
  Synergizing reasoning and acting in language models,'' in \emph{International
  Conference on Learning Representations (ICLR)}, 2023.

\bibitem{toolformer}
T.~Schick, J.~Dwivedi-Yu, R.~Dess{\`\i}, R.~Raileanu, M.~Lomeli, E.~Hambro,
  L.~Zettlemoyer, N.~Cancedda, and T.~Scialom, ``Toolformer: Language models
  can teach themselves to use tools,'' in \emph{Annual Conference on Neural
  Information Processing Systems (NeurIPS)}, 2023.

\bibitem{chatgpt-tools}
OpenAI, ``Chatgpt plugins,'' 2024,
  \url{https://openai.com/index/chatgpt-plugins/} (accessed 27, August, 2025).

\bibitem{microsoft-copilot}
Microsoft, ``What is microsoft 365 copilot?'' 2024,
  \url{https://learn.microsoft.com/en-us/microsoft-365/copilot/microsoft-365-copilot-overview}
  (accessed 27, August, 2025).

\bibitem{mcp}
{Model Context Protocol}, ``Model context protocol,'' 2025,
  \url{https://modelcontextprotocol.io/}.

\bibitem{openclaw}
{OpenClaw}, ``{OpenClaw},'' 2026, \url{https://openclaw.ai/}.

\bibitem{greshake2023not}
K.~Greshake, S.~Abdelnabi, S.~Mishra, C.~Endres, T.~Holz, and M.~Fritz, ``Not
  what you've signed up for: Compromising real-world llm-integrated
  applications with indirect prompt injection,'' in \emph{AISec Workshop},
  2023.

\bibitem{liu2024formalizing}
Y.~Liu, Y.~Jia, R.~Geng, J.~Jia, and N.~Z. Gong, ``Formalizing and benchmarking
  prompt injection attacks and defenses,'' in \emph{33rd USENIX Security
  Symposium}.\hskip 1em plus 0.5em minus 0.4em\relax USENIX Association, 2024.

\bibitem{injecagent}
\BIBentryALTinterwordspacing
Q.~Zhan, Z.~Liang, Z.~Ying, and D.~Kang, ``{InjecAgent}: Benchmarking indirect
  prompt injections in tool-integrated large language model agents,'' in
  \emph{Findings of the Association for Computational Linguistics: ACL
  2024}.\hskip 1em plus 0.5em minus 0.4em\relax Bangkok, Thailand: Association
  for Computational Linguistics, 2024, pp. 10\,471--10\,506. [Online].
  Available: \url{https://aclanthology.org/2024.findings-acl.624/}
\BIBentrySTDinterwordspacing

\bibitem{dualllm}
S.~Willison, ``The dual llm pattern for building ai assistants that can resist
  prompt injection,'' 2023,
  \url{https://simonwillison.net/2023/Apr/25/dual-llm-pattern/}.

\bibitem{pfi}
J.~Kim, W.~Choi, and B.~Lee, ``Prompt flow integrity to prevent privilege
  escalation in llm agents,'' \emph{arXiv preprint arXiv:2503.15547}, 2025.

\bibitem{camel}
E.~Debenedetti, I.~Shumailov, T.~Fan, J.~Hayes, N.~Carlini, D.~Fabian, C.~Kern,
  C.~Shi, A.~Terzis, and F.~Tramèr, ``Defeating prompt injections by design,''
  \emph{arXiv preprint arXiv:2503.18813}, 2025.

\bibitem{fides}
M.~Costa, B.~Köpf, A.~Kolluri, A.~Paverd, M.~Russinovich, A.~Salem, S.~Tople,
  L.~Wutschitz, and S.~Zanella-Béguelin, ``Securing ai agents with
  information-flow control,'' \emph{arXiv preprint arXiv:2505.23643}, 2025.

\bibitem{pinchbench}
{PinchBench}, ``{PinchBench},'' 2026, \url{https://pinchbench.com/}.

\bibitem{camels-computers}
H.~Foerster, T.~Blanchard, K.~Nikolić, I.~Shumailov, C.~Zhang, R.~Mullins,
  N.~Papernot, F.~Tramèr, and Y.~Zhao, ``{CaMeLs} can use computers too:
  System-level security for computer use agents,'' \emph{arXiv preprint
  arXiv:2601.09923}, 2026.

\bibitem{liu2024rce}
T.~Liu, Z.~Deng, G.~Meng, Y.~Li, and K.~Chen, ``Demystifying {RCE}
  vulnerabilities in {LLM}-integrated apps,'' in \emph{Proceedings of the 2024
  ACM SIGSAC Conference on Computer and Communications Security}.\hskip 1em
  plus 0.5em minus 0.4em\relax ACM, 2024.

\bibitem{asb}
\BIBentryALTinterwordspacing
H.~Zhang, J.~Huang, K.~Mei, Y.~Yao, Z.~Wang, C.~Zhan, H.~Wang, and Y.~Zhang,
  ``Agent security bench ({ASB}): Formalizing and benchmarking attacks and
  defenses in {LLM}-based agents,'' in \emph{13th International Conference on
  Learning Representations, ICLR 2025}.\hskip 1em plus 0.5em minus 0.4em\relax
  Singapore: International Conference on Learning Representations, 2025.
  [Online]. Available:
  \url{https://proceedings.iclr.cc/paper_files/paper/2025/hash/5750f91d8fb9d5c02bd8ad2c3b44456b-Abstract-Conference.html}
\BIBentrySTDinterwordspacing

\bibitem{agentvigil}
\BIBentryALTinterwordspacing
Z.~Wang, V.~Siu, Z.~Ye, T.~Shi, Y.~Nie, X.~Zhao, C.~Wang, W.~Guo, and D.~Song,
  ``{AGENTVIGIL}: Automatic black-box red-teaming for indirect prompt injection
  against {LLM} agents,'' in \emph{Findings of the Association for
  Computational Linguistics: EMNLP 2025}.\hskip 1em plus 0.5em minus
  0.4em\relax Suzhou, China: Association for Computational Linguistics, 2025,
  pp. 23\,159--23\,172. [Online]. Available:
  \url{https://aclanthology.org/2025.findings-emnlp.1258/}
\BIBentrySTDinterwordspacing

\bibitem{bipia}
\BIBentryALTinterwordspacing
J.~Yi, Y.~Xie, B.~Zhu, K.~Hines, E.~Kiciman, G.~Sun, X.~Xie, and F.~Wu,
  ``Benchmarking and defending against indirect prompt injection attacks on
  large language models,'' in \emph{Proceedings of the 31st ACM SIGKDD
  Conference on Knowledge Discovery and Data Mining}.\hskip 1em plus 0.5em
  minus 0.4em\relax ACM, 2025. [Online]. Available:
  \url{https://arxiv.org/abs/2312.14197}
\BIBentrySTDinterwordspacing

\bibitem{agentdojo}
E.~Debenedetti, J.~Zhang, M.~Balunovic, L.~Beurer-Kellner, M.~Fischer, and
  F.~Tram{\`e}r, ``{AgentDojo}: A dynamic environment to evaluate prompt
  injection attacks and defenses for {LLM} agents,'' in \emph{Annual Conference
  on Neural Information Processing Systems (NeurIPS) Datasets and Benchmarks
  Track}, 2024.

\bibitem{retrievalbarrier}
\BIBentryALTinterwordspacing
H.~Chang, E.~Bao, X.~Luo, and T.~Yu, ``Overcoming the retrieval barrier:
  Indirect prompt injection in the wild for {LLM} systems,'' in \emph{35th
  USENIX Security Symposium}.\hskip 1em plus 0.5em minus 0.4em\relax USENIX
  Association, 2026. [Online]. Available:
  \url{https://www.usenix.org/conference/usenixsecurity26/presentation/chang}
\BIBentrySTDinterwordspacing

\bibitem{obliinjection}
\BIBentryALTinterwordspacing
R.~Wang, Y.~Jia, and N.~Z. Gong, ``{ObliInjection}: Order-oblivious prompt
  injection attack to {LLM} agents with multi-source data,'' in \emph{Network
  and Distributed System Security Symposium 2026}.\hskip 1em plus 0.5em minus
  0.4em\relax San Diego, CA, USA: Internet Society, 2026. [Online]. Available:
  \url{https://www.ndss-symposium.org/ndss-paper/obliinjection-order-oblivious-prompt-injection-attack-to-llm-agents-with-multi-source-data/}
\BIBentrySTDinterwordspacing

\bibitem{lesdissonances}
\BIBentryALTinterwordspacing
Z.~Li, J.~Cui, X.~Liao, and L.~Xing, ``Les dissonances: Cross-tool harvesting
  and polluting in pool-of-tools empowered {LLM} agents,'' in \emph{Network and
  Distributed System Security Symposium 2026}.\hskip 1em plus 0.5em minus
  0.4em\relax San Diego, CA, USA: Internet Society, 2026. [Online]. Available:
  \url{https://www.ndss-symposium.org/ndss-paper/les-dissonances-cross-tool-harvesting-and-polluting-in-pool-of-tools-empowered-llm-agents/}
\BIBentrySTDinterwordspacing

\bibitem{toolhijacker}
\BIBentryALTinterwordspacing
J.~Shi, Z.~Yuan, G.~Tie, P.~Zhou, N.~Z. Gong, and L.~Sun, ``Prompt injection
  attack to tool selection in {LLM} agents,'' in \emph{Network and Distributed
  System Security Symposium 2026}.\hskip 1em plus 0.5em minus 0.4em\relax San
  Diego, CA, USA: Internet Society, 2026. [Online]. Available:
  \url{https://www.ndss-symposium.org/ndss-paper/prompt-injection-attack-to-tool-selection-in-llm-agents/}
\BIBentrySTDinterwordspacing

\bibitem{spaiware}
\BIBentryALTinterwordspacing
M.~Herrador and J.~Rehberger, ``{SpAIware}: Uncovering a novel artificial
  intelligence attack vector through persistent memory in {LLM} applications
  and agents,'' \emph{Future Generation Computer Systems}, vol. 174, p. 107994,
  2026. [Online]. Available:
  \url{https://www.sciencedirect.com/science/article/pii/S0167739X25002894}
\BIBentrySTDinterwordspacing

\bibitem{minja}
\BIBentryALTinterwordspacing
S.~Dong, S.~Xu, P.~He, Y.~Li, J.~Tang, T.~Liu, H.~Liu, and Z.~Xiang, ``Memory
  injection attacks on {LLM} agents via query-only interaction,'' in \emph{ICLR
  2026 Workshop on Memory for {LLM}-Based Agentic Systems (MemAgents)}, 2026.
  [Online]. Available: \url{https://openreview.net/forum?id=i7J62t2wtV}
\BIBentrySTDinterwordspacing

\bibitem{promptinfection}
\BIBentryALTinterwordspacing
D.~Lee and M.~Tiwari, ``Prompt infection: {LLM}-to-{LLM} prompt injection
  within multi-agent systems,'' \emph{arXiv preprint arXiv:2410.07283}, 2024.
  [Online]. Available: \url{https://arxiv.org/abs/2410.07283}
\BIBentrySTDinterwordspacing

\bibitem{airgapagent}
\BIBentryALTinterwordspacing
E.~Bagdasarian, R.~Yi, S.~Ghalebikesabi, P.~Kairouz, M.~Gruteser, S.~Oh,
  B.~Balle, and D.~Ramage, ``{AirGapAgent}: Protecting privacy-conscious
  conversational agents,'' in \emph{Proceedings of the 2024 ACM SIGSAC
  Conference on Computer and Communications Security}.\hskip 1em plus 0.5em
  minus 0.4em\relax ACM, 2024. [Online]. Available:
  \url{https://dl.acm.org/doi/10.1145/3658644.3690350}
\BIBentrySTDinterwordspacing

\bibitem{fsecure}
F.~Wu, E.~Cecchetti, and C.~Xiao, ``System-level defense against indirect
  prompt injection attacks: An information flow control perspective,''
  \emph{arXiv preprint arXiv:2409.19091}, 2024.

\bibitem{ace}
\BIBentryALTinterwordspacing
E.~Li, T.~Mallick, E.~Rose, W.~Robertson, A.~Oprea, and C.~Nita-Rotaru,
  ``{ACE}: A security architecture for {LLM}-integrated app systems,'' in
  \emph{Network and Distributed System Security Symposium 2026}.\hskip 1em plus
  0.5em minus 0.4em\relax San Diego, CA, USA: Internet Society, 2026. [Online].
  Available:
  \url{https://www.ndss-symposium.org/ndss-paper/ace-a-security-architecture-for-llm-integrated-app-systems/}
\BIBentrySTDinterwordspacing

\bibitem{ignorepreviousprompt}
\BIBentryALTinterwordspacing
F.~Perez and I.~Ribeiro, ``Ignore previous prompt: Attack techniques for
  language models,'' \emph{arXiv preprint arXiv:2211.09527}, 2022. [Online].
  Available: \url{https://arxiv.org/abs/2211.09527}
\BIBentrySTDinterwordspacing

\bibitem{jailbroken}
\BIBentryALTinterwordspacing
A.~Wei, N.~Haghtalab, and J.~Steinhardt, ``Jailbroken: How does {LLM} safety
  training fail?'' \emph{arXiv preprint arXiv:2307.02483}, 2023. [Online].
  Available: \url{https://arxiv.org/abs/2307.02483}
\BIBentrySTDinterwordspacing

\bibitem{zou2023universal}
\BIBentryALTinterwordspacing
A.~Zou, Z.~Wang, N.~Carlini, M.~Nasr, J.~Z. Kolter, and M.~Fredrikson,
  ``Universal and transferable adversarial attacks on aligned language
  models,'' \emph{arXiv preprint arXiv:2307.15043}, 2023. [Online]. Available:
  \url{https://arxiv.org/abs/2307.15043}
\BIBentrySTDinterwordspacing

\bibitem{instructhier}
E.~Wallace, K.~Xiao, R.~Leike, L.~Weng, J.~Heidecke, and A.~Beutel, ``The
  instruction hierarchy: Training {LLMs} to prioritize privileged
  instructions,'' \emph{arXiv preprint arXiv:2404.13208}, 2024.

\bibitem{struq}
\BIBentryALTinterwordspacing
S.~Chen, J.~Piet, C.~Sitawarin, and D.~Wagner, ``{StruQ}: Defending against
  prompt injection with structured queries,'' in \emph{34th USENIX Security
  Symposium}.\hskip 1em plus 0.5em minus 0.4em\relax USENIX Association, 2025.
  [Online]. Available: \url{https://arxiv.org/abs/2402.06363}
\BIBentrySTDinterwordspacing

\bibitem{secalign}
\BIBentryALTinterwordspacing
S.~Chen, A.~Zharmagambetov, S.~Mahloujifar, K.~Chaudhuri, D.~Wagner, and
  C.~Guo, ``{SecAlign}: Defending against prompt injection with preference
  optimization,'' in \emph{Proceedings of the 2025 ACM SIGSAC Conference on
  Computer and Communications Security}.\hskip 1em plus 0.5em minus 0.4em\relax
  ACM, 2025. [Online]. Available:
  \url{https://dl.acm.org/doi/10.1145/3719027.3744836}
\BIBentrySTDinterwordspacing

\bibitem{promptguard}
{Meta AI}, ``{Llama Prompt Guard 2},'' 2025,
  \url{https://huggingface.co/meta-llama/Llama-Prompt-Guard-2-86M}.

\bibitem{llamafirewall}
S.~Chennabasappa, C.~Nikolaidis, D.~Song, D.~Molnar, S.~Ding, S.~Wan,
  S.~Whitman, L.~Deason, N.~Doucette, A.~Montilla, A.~Gampa, B.~de~Paola,
  D.~Gabi, J.~Crnkovich, J.-C. Testud, K.~He, R.~Chaturvedi, W.~Zhou, and
  J.~Saxe, ``{LlamaFirewall}: An open source guardrail system for building
  secure {AI} agents,'' \emph{arXiv preprint arXiv:2505.03574}, 2025,
  \url{https://github.com/meta-llama/PurpleLlama/tree/main/LlamaFirewall}.

\bibitem{datasentinel}
\BIBentryALTinterwordspacing
Y.~Liu, Y.~Jia, J.~Jia, D.~Song, and N.~Z. Gong, ``{DataSentinel}: A
  game-theoretic detection of prompt injection attacks,'' in \emph{IEEE
  Symposium on Security and Privacy}.\hskip 1em plus 0.5em minus 0.4em\relax
  IEEE, 2025. [Online]. Available: \url{https://arxiv.org/abs/2504.11358}
\BIBentrySTDinterwordspacing

\bibitem{spotlighting}
K.~Hines, G.~Lopez, M.~Hall, F.~Zarfati, Y.~Zunger, and E.~Kiciman, ``Defending
  against indirect prompt injection attacks with spotlighting,'' \emph{arXiv
  preprint arXiv:2403.14720}, 2024.

\bibitem{rennervate}
\BIBentryALTinterwordspacing
Y.~Zhong, Q.~Miao, Y.~Chen, J.~Deng, Y.~Cheng, and W.~Xu, ``Attention is all
  you need to defend against indirect prompt injection attacks in {LLMs},'' in
  \emph{Network and Distributed System Security Symposium 2026}.\hskip 1em plus
  0.5em minus 0.4em\relax San Diego, CA, USA: Internet Society, 2026. [Online].
  Available:
  \url{https://www.ndss-symposium.org/ndss-paper/attention-is-all-you-need-to-defend-against-indirect-prompt-injection-attacks-in-llms/}
\BIBentrySTDinterwordspacing

\bibitem{melon}
\BIBentryALTinterwordspacing
K.~Zhu, X.~Yang, J.~Wang, W.~Guo, and W.~Y. Wang, ``{MELON}: Provable defense
  against indirect prompt injection attacks in {AI} agents,'' in
  \emph{Proceedings of the 42nd International Conference on Machine
  Learning}.\hskip 1em plus 0.5em minus 0.4em\relax PMLR, 2025. [Online].
  Available: \url{https://arxiv.org/abs/2502.05174}
\BIBentrySTDinterwordspacing

\bibitem{taskshield}
\BIBentryALTinterwordspacing
F.~Jia, T.~Wu, X.~Qin, and A.~Squicciarini, ``The task shield: Enforcing task
  alignment to defend against indirect prompt injection in {LLM} agents,'' in
  \emph{Proceedings of the 63rd Annual Meeting of the Association for
  Computational Linguistics (Volume 1: Long Papers)}.\hskip 1em plus 0.5em
  minus 0.4em\relax Association for Computational Linguistics, 2025. [Online].
  Available: \url{https://aclanthology.org/2025.acl-long.1435/}
\BIBentrySTDinterwordspacing

\bibitem{driftdef}
\BIBentryALTinterwordspacing
H.~Li, X.~Liu, H.-C. Chiu, D.~Li, N.~Zhang, and C.~Xiao, ``{DRIFT}: Dynamic
  rule-based defense with injection isolation for securing {LLM} agents,'' in
  \emph{Annual Conference on Neural Information Processing Systems (NeurIPS)},
  2025. [Online]. Available: \url{https://arxiv.org/abs/2506.12104}
\BIBentrySTDinterwordspacing

\bibitem{shieldagent}
\BIBentryALTinterwordspacing
Z.~Chen, M.~Kang, and B.~Li, ``{ShieldAgent}: Shielding agents via verifiable
  safety policy reasoning,'' in \emph{Proceedings of the 42nd International
  Conference on Machine Learning}.\hskip 1em plus 0.5em minus 0.4em\relax PMLR,
  2025. [Online]. Available: \url{https://arxiv.org/abs/2503.22738}
\BIBentrySTDinterwordspacing

\bibitem{guardagent}
\BIBentryALTinterwordspacing
Z.~Xiang, L.~Zheng, Y.~Li, J.~Hong, Q.~Li, H.~Xie, J.~Zhang, Z.~Xiong, C.~Xie,
  C.~Yang, D.~Song, and B.~Li, ``{GuardAgent}: Safeguard {LLM} agents by a
  guard agent via knowledge-enabled reasoning,'' in \emph{Proceedings of the
  42nd International Conference on Machine Learning}.\hskip 1em plus 0.5em
  minus 0.4em\relax PMLR, 2025. [Online]. Available:
  \url{https://arxiv.org/abs/2406.09187}
\BIBentrySTDinterwordspacing

\bibitem{agentspec}
\BIBentryALTinterwordspacing
H.~Wang, C.~M. Poskitt, and J.~Sun, ``{AgentSpec}: Customizable runtime
  enforcement for safe and reliable {LLM} agents,'' in \emph{Proceedings of the
  48th IEEE/ACM International Conference on Software Engineering}.\hskip 1em
  plus 0.5em minus 0.4em\relax ACM, 2026. [Online]. Available:
  \url{https://arxiv.org/abs/2503.18666}
\BIBentrySTDinterwordspacing

\bibitem{rtbas}
P.~Y. Zhong, S.~Chen, R.~Wang, M.~McCall, B.~L. Titzer, H.~Miller, and P.~B.
  Gibbons, ``{RTBAS}: Defending {LLM} agents against prompt injection and
  privacy leakage,'' \emph{arXiv preprint arXiv:2502.08966}, 2025.

\bibitem{permissiveifc}
S.~A. Siddiqui, R.~Gaonkar, B.~Köpf, D.~Krueger, A.~Paverd, A.~Salem,
  S.~Tople, L.~Wutschitz, M.~Xia, and S.~Zanella-Béguelin, ``Permissive
  information-flow analysis for large language models,'' \emph{arXiv preprint
  arXiv:2410.03055}, 2024.

\bibitem{judgedeceiver}
J.~Shi, Z.~Yuan, Y.~Liu, Y.~Huang, P.~Zhou, L.~Sun, and N.~Z. Gong,
  ``Optimization-based prompt injection attack to {LLM}-as-a-judge,'' in
  \emph{Proceedings of the 2024 ACM SIGSAC Conference on Computer and
  Communications Security}.\hskip 1em plus 0.5em minus 0.4em\relax ACM, 2024.

\bibitem{openshell}
{NVIDIA}, ``{OpenShell}: A safe, private runtime for autonomous {AI} agents,''
  2026, \url{https://github.com/NVIDIA/OpenShell}.

\bibitem{openclaw-webhooks}
{OpenClaw}, ``{Webhooks},'' 2026, \url{https://docs.openclaw.ai/cli/webhooks}.

\bibitem{openclaw-configuration}
------, ``{Configuration},'' 2026,
  \url{https://docs.openclaw.ai/gateway/configuration} (accessed 29, May,
  2026).

\bibitem{git-worktree}
{Git Project}, \emph{{git-worktree Documentation}}, 2026,
  \url{https://git-scm.com/docs/git-worktree} (accessed 28, May, 2026).

\bibitem{gws}
{Google Workspace}, ``{gws}: {Google Workspace} {CLI},'' 2026,
  \url{https://github.com/googleworkspace/cli} (accessed 5, June, 2026).

\bibitem{gh}
{GitHub}, ``{GitHub CLI},'' 2026, \url{https://cli.github.com/} (accessed 5,
  June, 2026).

\bibitem{autodojo}
X.~Ma, T.~Li, C.~Xiao, Z.~Yu, N.~Zhang, and Y.~Vorobeychik, ``{AutoDojo}:
  Adaptive black-box attacks reveal the limits of {IPI} defenses and
  task-specification effects in {LLM} agents,'' \emph{arXiv preprint
  arXiv:2606.15057}, 2026.

\bibitem{imprompter-attack}
\BIBentryALTinterwordspacing
X.~Fu, S.~Li, Z.~Wang, Y.~Liu, R.~K. Gupta, T.~Berg-Kirkpatrick, and
  E.~Fernandes, ``{Imprompter}: Tricking {LLM} agents into improper tool use,''
  \emph{arXiv preprint arXiv:2410.14923}, 2024. [Online]. Available:
  \url{https://arxiv.org/abs/2410.14923}
\BIBentrySTDinterwordspacing

\bibitem{crescendo}
M.~Russinovich, A.~Salem, and R.~Eldan, ``Great, now write an article about
  that: The crescendo multi-turn {LLM} jailbreak attack,'' in \emph{34th USENIX
  Security Symposium}.\hskip 1em plus 0.5em minus 0.4em\relax USENIX
  Association, 2025.

\bibitem{poisonedrag}
W.~Zou, R.~Geng, B.~Wang, and J.~Jia, ``{PoisonedRAG}: Knowledge corruption
  attacks to retrieval-augmented generation of large language models,'' in
  \emph{34th USENIX Security Symposium}.\hskip 1em plus 0.5em minus 0.4em\relax
  USENIX Association, 2025.

\bibitem{secgpt}
\BIBentryALTinterwordspacing
Y.~Wu, F.~Roesner, T.~Kohno, N.~Zhang, and U.~Iqbal, ``An execution isolation
  architecture for {LLM}-based agentic systems,'' in \emph{Network and
  Distributed System Security Symposium 2025}.\hskip 1em plus 0.5em minus
  0.4em\relax San Diego, CA, USA: Internet Society, 2025. [Online]. Available:
  \url{https://arxiv.org/abs/2403.04960}
\BIBentrySTDinterwordspacing

\bibitem{prudentia}
A.~Kolluri, R.~Sharma, M.~Costa, B.~Köpf, T.~Nießen, M.~Russinovich,
  S.~Tople, and S.~Zanella-Béguelin, ``Optimizing agent planning for security
  and autonomy,'' in \emph{International Conference on Learning Representations
  (ICLR)}, 2026.

\bibitem{progent}
T.~Shi, J.~He, Z.~Wang, H.~Li, L.~Wu, W.~Guo, and D.~Song, ``{Progent}:
  Programmable privilege control for {LLM} agents,'' \emph{arXiv preprint
  arXiv:2504.11703}, 2025.

\bibitem{conseca}
\BIBentryALTinterwordspacing
L.~Tsai and E.~Bagdasarian, ``{Conseca}: Contextual agent security: A policy
  for every purpose,'' in \emph{Proceedings of the 20th Workshop on Hot Topics
  in Operating Systems}.\hskip 1em plus 0.5em minus 0.4em\relax ACM, 2025.
  [Online]. Available: \url{https://arxiv.org/abs/2501.17070}
\BIBentrySTDinterwordspacing

\bibitem{saga}
G.~Syros, A.~Suri, J.~Ginesin, C.~Nita-Rotaru, and A.~Oprea, ``{SAGA}: A
  security architecture for governing {AI} agentic systems,'' in \emph{Network
  and Distributed System Security Symposium 2026}.\hskip 1em plus 0.5em minus
  0.4em\relax San Diego, CA, USA: Internet Society, 2026.

\bibitem{claude-code-hooks}
{Anthropic}, ``{Claude Code} hooks reference,'' 2026,
  \url{https://code.claude.com/docs/en/hooks#hooks-reference} (accessed 11,
  June, 2026).

\bibitem{hermes-agent-hooks}
{Nous Research}, ``{Hermes Agent}: Hooks,'' 2026,
  \url{https://hermes-agent.nousresearch.com/docs/user-guide/features/hooks}
  (accessed 11, June, 2026).

\bibitem{liu2025make}
F.~Liu, Y.~Zhang, J.~Luo, J.~Dai, T.~Chen, L.~Yuan, Z.~Yu, Y.~Shi, K.~Li,
  C.~Zhou \emph{et~al.}, ``Make agent defeat agent: Automatic detection of
  {Taint-Style} vulnerabilities in {LLM-based} agents,'' in \emph{34th USENIX
  Security Symposium}.\hskip 1em plus 0.5em minus 0.4em\relax USENIX
  Association, 2025, pp. 3767--3786.

\bibitem{langchain-middleware}
{LangChain}, ``{LangChain} middleware,'' 2026,
  \url{https://docs.langchain.com/oss/javascript/langchain/middleware}
  (accessed 30, August, 2026).

\bibitem{ndss-2026}
\emph{Network and Distributed System Security Symposium 2026}.\hskip 1em plus
  0.5em minus 0.4em\relax San Diego, CA, USA: Internet Society, 2026.

\bibitem{icml-2025}
\emph{Proceedings of the 42nd International Conference on Machine
  Learning}.\hskip 1em plus 0.5em minus 0.4em\relax PMLR, 2025.

\bibitem{ccs-2024}
\emph{Proceedings of the 2024 ACM SIGSAC Conference on Computer and
  Communications Security}.\hskip 1em plus 0.5em minus 0.4em\relax ACM, 2024.

\bibitem{usenix-security-2025}
\emph{34th USENIX Security Symposium}.\hskip 1em plus 0.5em minus 0.4em\relax
  USENIX Association, 2025.

\end{thebibliography}
}
\newpage
\appendices

\section{Implementation}
\label{s:appendix:implementation}
%

\PP{OpenClaw integration}
\sys is implemented with event hooks and custom tools, a mechanism
supported in many personal agents~(\eg Claude
Code~\cite{claude-code-hooks} and Hermes
Agent~\cite{hermes-agent-hooks}).
This hook-based design does not require changing the AI model, the agent
main logic, or tool implementations, which makes it deployable on
various personal agents.
As in prior Dual LLM pattern defenses~\cite{pfi,fides,camel}, \sys also
registers \cc{inspect\_symbol} as a custom tool for \ULLM processing, which
takes the symbols, the processing instruction for \ULLM, and an output
schema for the result.
\sys checks the \ULLM result against the schema, returns the symbolized
result to \AgentView when it matches, and otherwise returns an error without
exposing any untrusted data.

\begin{figure}[t]
\centering
\begin{tikzpicture}[
  font=\scriptsize,
  event/.style={draw, rounded corners, fill=blue!8, align=center,
    text width=1.3cm, minimum height=0.5cm, inner sep=1.5pt},
  hook/.style={draw, fill=black!6, font=\ttfamily\tiny,
    align=center, inner sep=2pt},
  action/.style={draw, rounded corners=2pt, fill=yellow!14, align=left,
    text width=3.4cm, inner sep=2.5pt, font=\scriptsize},
  flow/.style={-{Latex[length=1.4mm]}, thick},
  link/.style={-{Latex[length=1.2mm]}},
  loop/.style={-{Latex[length=1.4mm]}, dashed, gray!70},
]
\node[font=\bfseries\scriptsize] at (0.15,0.75) {Agent event};
\node[font=\bfseries\scriptsize] at (2.3,0.75) {OpenClaw hook};
\node[font=\bfseries\scriptsize] at (5.4,0.75) {\sys action};

\node[event] (e1) at (0.15, 0)    {Agent initialization};
\node[event] (e2) at (0.15,-1.5)  {Receive user prompt};
\node[event] (e3) at (0.15,-3.0)  {Tool call decision};
\node[event] (e4) at (0.15,-4.5)  {Tool call};
\node[event] (e5) at (0.15,-6.0)  {Tool result returns};
\node[event] (e6) at (0.15,-7.5)  {Return response};
\draw[flow] (e1) -- (e2);
\draw[flow] (e2) -- (e3);
\draw[flow] (e3) -- (e4);
\draw[flow] (e4) -- (e5);
\draw[flow] (e5) -- (e6);
\draw[loop] (e5.west) .. controls +(left:0.65) and +(left:0.65) ..
  node[midway,font=\scriptsize,rotate=90,fill=white,inner sep=1pt,text=gray!85]{Agent Loop} (e3.west);

\node[hook] (h1) at (2.3, 0)    {before\_prompt\_build};
\node[hook] (h2) at (2.3,-3.0)  {before\_tool\_call};
\node[hook] (h3) at (2.3,-5.6)  {after\_tool\_call};
\node[hook] (h4) at (2.3,-6.5)  {transform\_tool\_result$^\dagger$};
\node[hook] (h5) at (2.3,-7.5)  {message\_sending$^\dagger$};

\node[action] (a1) at (5.4, 0)    {load policy and symbol table, add system prompt that instructs the agent how to handle symbols};
\node[action] (a2) at (5.4,-2.55) {\AgentView tool: copy \HumanView file changes into \AgentView};
\node[action] (a3) at (5.4,-3.5)  {\HumanView tool: check data usage policy, resolve needed symbols};
\node[action] (a4) at (5.4,-5.6)  {\AgentView tool: copy \AgentView file changes into \HumanView};
\node[action] (a5) at (5.4,-6.5)  {\HumanView tool: symbolize untrusted data based on data trust policy};
\node[action] (a6) at (5.4,-7.5)  {Agent response: resolve symbols};

\draw[link] (e1.east) -- (h1.west);
\draw[link] (e3.east) -- (h2.west);
\draw[link] (e5.east) -- (h3.west);
\draw[link] (e5.east) -- (h4.west);
\draw[link] (e6.east) -- (h5.west);

\draw[link] (h1.east) -- (a1.west);
\draw[link] (h2.east) -- (a2.west);
\draw[link] (h2.east) -- (a3.west);
\draw[link] (h3.east) -- (a4.west);
\draw[link] (h4.east) -- (a5.west);
\draw[link] (h5.east) -- (a6.west);
\end{tikzpicture}
\caption{OpenClaw hook points that \sys uses across the agent's events.
\cc{before\_tool\_call} and the two result hooks branch by whether the tool
runs on \AgentView or \HumanView. \sys changes the two hooks marked
$\dagger$.}
\label{fig:appendix-hooks}
\end{figure}
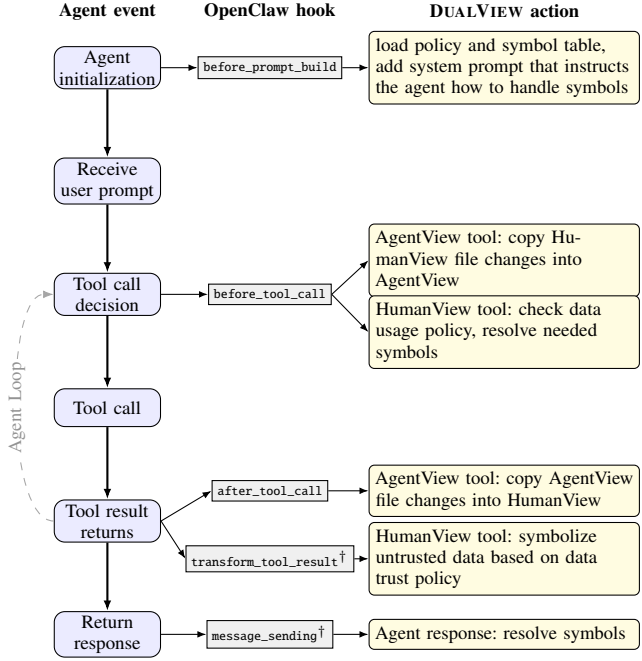

\PP{Hook points}
OpenClaw hooks let \sys mediate the agent lifecycle at initialization,
tool execution, and human-facing replies~(\autoref{fig:appendix-hooks}).
The pre-tool hook handles the work after \TLLM selects a tool but before
the tool runs: for \AgentView tools it copies \HumanView file changes
into the \AgentFileSystem, and for \HumanView tools it checks the
data usage policy and resolves the symbols the tool needs as
original data.
The post-tool result hook applies only to \HumanView tools, which can
return original untrusted data that must be symbolized before \TLLM
reads it.
\sys uses its \cc{transform\_tool\_result} hook to apply the data trust
policy to web results, \HumanShell stdout, webhook payloads, and other
\HumanView tool results before \TLLM reads them; OpenClaw v2026.3.12 did
not support this hook, so \sys added it, and the latest OpenClaw
provides it natively as tool-result middleware.
Final assistant messages use a separate delivery hook, since no tool
call occurs after \TLLM generates the final text.
\sys uses \cc{message\_sending} to resolve symbols only in the agent
response that goes to the CLI, webchat, or channel, so the transcript
keeps symbols while the human-facing message contains original data;
OpenClaw v2026.3.12 did not deliver an edited response to the messenger
channel, so \sys patched it.

%

\PP{Symbol table}
\sys stores the symbol table in SQLite and shares it across all agent
sessions.
The symbol table maps each symbol to its original data and its
provenance, allowing \sys to resolve symbols when a symbol is passed
to \HumanView.
\sys resolves a symbol only on an exact match of its full token, so a
command or edit that splits, truncates, or otherwise mangles a symbol
leaves a fragment that \sys does not resolve.
Such a fragment never reveals the original untrusted data, so resolution
fails closed; it can leave a broken token in human-facing output and
lower human utility, a case we did not observe in our evaluation.
Each entry stores a symbol with the original data and provenance
metadata.
The provenance metadata comes from trusted data~(\eg the originating tool,
the field path, and the data origin).
\sys can extend the data usage policy with this provenance in the
future, for example to block symbols of a given origin from sensitive tool
arguments, which prevents taint-style vulnerabilities~\cite{liu2025make}.
%
%

\PP{System Prompt}
\sys appends a system prompt to agent context that explains how to
handle its symbolized tool results and shell tools.
It instructs the model to treat symbol tokens as opaque handles and pass
them through unchanged, since they are resolved to their real values only at
delivery time, and to call the \cc{inspect\_symbol} tool whenever it needs
a derived result such as a summary, an extraction, or a transformation of
symbolized data.
It also documents the Data Trust Policy together with its
\cc{policy\_add} and \cc{policy\_del} tools, and the two shell
modes~(\ie shell tools).
Default exec resolves symbols and marks the output untrusted, while
restricted exec (\cc{RESTRICTED=1}) runs sandboxed and returns trusted
output.
The full prompt text is in our released code.

\PP{Filesystem}
\sys tracks the files the agent modifies with Git and keeps the
\AgentFileSystem and \HumanFileSystem as two Git
worktrees~\cite{git-worktree}~(\autoref{s:design:file-tools}).
\autoref{fig:appendix-fs-layout} shows an example layout of file systems.
The original workspace directory is the \HumanFileSystem, and \sys keeps
its own Git database and the \AgentFileSystem outside that directory under
\cc{\textasciitilde/.dualview/workspaces/}, where it canonicalizes
and percent-encodes the root path into a single directory name.
Because \sys keeps its Git database outside the workspace, its Git use stays
transparent to the user, so ordinary Git commands in the workspace work
seamlessly.
\sys starts tracking a new workspace on demand when the agent first writes a
file with a tool such as \cc{write} or \cc{edit}.
If the file is in a Git project, the workspace root is the enclosing project
directory, otherwise it is the parent directory of the file.
Unlike a file tool, a shell command can write to paths \sys cannot
anticipate, so \sys does not open a new workspace for \AgentShell and
instead confines its writes to the tracked workspaces, keeping every
write inside a worktree that the post-tool sync reconciles.

\sys handles potential file write race conditions in two ways.
First, a race between two agents that sync the same root is handled by
serializing sync commits with a workspace-scoped file lock.
Second, a race between a human write and a sync, which the lock does not
prevent, is handled by detecting the changed file hash, skipping the
affected file so \sys never overwrites the human's change, and appending a
record to the conflict log.
\sys currently only reports the conflict through the log and leaves the
resolution to the human.

We evaluated file-system synchronization under concurrent access
in \autoref{s:eval:concurrency}.

\begin{figure}[t]
\centering
\scriptsize
\begin{tabular}{@{}ll@{}}
\toprule
\cc{/home/user/apple/banana/} & workspace \\
\quad ordinary workspace files & \emph{\HumanView files} \\
\quad \cc{.git/} & ordinary Git, optional \\
\addlinespace
\cc{\textasciitilde/.dualview/workspaces/} & \sys file-system management \\
\quad \cc{\%2Fhome\%2Fuser\%2Fapple\%2Fbanana/} & encoded workspace path \\
\quad\quad \cc{repo.git/} & \sys Git \\
\quad\quad \cc{agentview/} & \emph{\AgentFileSystem} for the workspace \\
\quad\quad \cc{conflicts.log} & sync conflict log \\
\bottomrule
\end{tabular}
\caption{Example file-system layout under \sys.}
\label{fig:appendix-fs-layout}
\end{figure}

\PP{Claude Code integration}

We implemented a prototype of \sys as a Claude Code plugin using
hooks and additional MCP tools~\cite{claude-code-hooks}.
The \cc{PreToolUse} and \cc{PostToolUse} hooks mediate the inputs and
outputs of native \cc{Read} and \cc{Write}, \cc{SessionStart} adds
symbol instructions, and \cc{MessageDisplay} resolves user responses.
Additional MCP tools provide editing, shell execution, web fetch, and
\ULLM processing~\cite{mcp}.
The plugin reuses \sys's policies, file-system views, shell modes, and
synchronization.

\PP{LangChain integration}

We implemented a prototype of \sys as LangChain middleware around
model and tool calls~\cite{langchain-middleware}.
The middleware adds symbol instructions before model calls, mediates
tool inputs and outputs before results enter the message history, and
resolves final responses.
It reuses \sys's policies, file-system views, shell modes, and
synchronization without changing LangChain's agent loop.

\section{\sys Policy}
\label{s:appendix:policy}

\begin{table}[t]
\centering
\footnotesize
\caption{PinchBench file paths marked as untrusted. These patterns match
  49 of 98 input file paths, and 42 of the 147
  PinchBench tasks read at least one of them.}
\label{tab:pinchbench-untrusted-files}
\begin{tabular*}{\columnwidth}{@{\extracolsep{\fill}}>{\raggedright\arraybackslash}p{0.26\columnwidth}
                  >{\raggedright\arraybackslash}p{0.62\columnwidth}@{}}
\toprule
\textbf{Category} & \textbf{Untrusted Files} \\
\midrule
Email and inbox &
\cc{emails/*}, \cc{inbox/*} \\
\addlinespace
External documents and reports &
\cc{ai\_blog.txt}, \cc{GPT4.pdf}, \cc{openclaw\_report.pdf},
\cc{research/*}, \cc{sample\_contract.pdf},
\cc{school-calendar.pdf}, \cc{subway\_map.md},
\cc{vulnerability\_scan.json} \\
\addlinespace
Logs and request metadata &
\cc{access\_events.csv}, \cc{apache\_error.log}, \cc{auth.log},
\cc{hdfs\_datanode.log}, \cc{linux\_syslog.log},
\cc{mapreduce.log}, \cc{nginx\_access.log}, \cc{syslog.log} \\
\bottomrule
\end{tabular*}
\end{table}

\begin{table}[t]
\centering
\footnotesize
\caption{Sensitive PinchBench output directories made read-only for
  the OpenShell utility evaluation. 60 of 147 tasks write to these
  directories.}
\label{tab:pinchbench-output-routes}
\begin{tabular*}{\columnwidth}{@{\extracolsep{\fill}}>{\raggedright\arraybackslash}p{0.52\columnwidth}
                  >{\raggedright\arraybackslash}p{0.38\columnwidth}@{}}
\toprule
\textbf{Category} & \textbf{Restricted Path} \\
\midrule
Email replies & \cc{output/email\_replies/} \\
Security, contract, and triage reports & \cc{output/security/} \\
Meeting decisions, recommendations, and summaries & \cc{output/decisions/} \\
CI, Kubernetes, logs, and service actions & \cc{output/operations/} \\
Research, finance, procurement, and stock outputs & \cc{output/business/} \\
\bottomrule
\end{tabular*}
\end{table}

\PP{Policy specification}
\autoref{tab:appendix-data-trust-policy} summarizes the prototype data
trust policy that \sys uses for OpenClaw built-in tools and webhook
receivers.
Fields listed as trusted pass through to the agent as original data.
Fields listed as untrusted are symbolized before the agent reads them.
Webhook fields omitted from an endpoint schema are untrusted by default.

\begin{table*}[t]
\centering
\scriptsize
\caption{Prototype data trust policy specification. The policy uses the same
  trusted and untrusted field distinction for built-in tools, structured
  shell output, webhook receivers, and inter-agent communication.}
\label{tab:appendix-data-trust-policy}
\begin{tabular}{@{}p{0.16\textwidth}p{0.39\textwidth}p{0.37\textwidth}@{}}
\toprule
\textbf{Tool or source} & \textbf{Trusted fields} & \textbf{Untrusted fields} \\
\midrule
\cc{web\_fetch} &
\cc{url}, \cc{status}, \cc{contentType},
\cc{extractMode}, \cc{extractor}, \cc{fetchedAt}, \cc{tookMs},
\cc{truncated}, \cc{length}, \cc{rawLength}, \cc{wrappedLength},
\cc{externalContent}, \cc{warning} &
\cc{finalUrl}, \cc{title}, \cc{text} \\
\addlinespace
\cc{web\_search} &
\cc{query}, \cc{provider}, \cc{count}, \cc{tookMs},
\cc{externalContent} &
Per-result \cc{title}, \cc{description}, \cc{url}, \cc{published},
\cc{siteName} \\
\addlinespace
\cc{exec} &
Fields declared by per-skill output schemas, such as local counters,
identifiers, and schema tags &
Default stdout, plus remote or user-controlled fields declared by
per-skill output schemas \\
\addlinespace
\cc{read} &
File content from \AgentView, where untrusted content is already
symbolized &
None \\
\addlinespace
\cc{write} and \cc{edit} &
Tool status and other local metadata &
None \\
\addlinespace
\cc{process} &
Write and lifecycle actions, \cc{list}, and trusted local metadata &
Read actions such as \cc{poll} and \cc{log}, plus unknown actions \\
\addlinespace
\cc{inspect\_symbol} &
Validated output after \ULLM processing and re-symbolization &
None \\
\addlinespace
Webhook receivers &
Endpoint-specific schema fields such as verified source metadata,
event type, and local delivery identifiers &
Endpoint-specific schema fields such as message text, issue comments,
form bodies, and fields omitted from the schema \\
\addlinespace
Image and media tools &
Local tool metadata and URL-keyed identifiers &
Remote captions, descriptions, or other service-controlled text when
present \\
\addlinespace
Inter-agent session tools &
Local conversation identifiers and private context metadata &
Message text and tool-result text from untrusted agent contexts \\
\addlinespace
Other local built-in tools &
\cc{memory\_search}, \cc{memory\_get}, \cc{session\_status},
\cc{canvas}, \cc{nodes}, \cc{cron}, \cc{tts}, \cc{message}, and
\cc{claude\_code\_*} results that contain local metadata only &
Remote or externally controlled text when a deployment configures such
fields \\
\bottomrule
\end{tabular}
\end{table*}

\PP{Examples}
\autoref{tab:policy-rule-examples} shows how the two \sys policies
apply at different boundaries.
The upper block applies the data trust policy after a \HumanView tool
returns.
The rule on the left names the trusted and untrusted fields or origins,
and the result on the right is what \TLLM sees after \sys replaces
untrusted fields with symbols.
The lower block applies the data usage policy before
\HumanShell runs a command.
Commands with no symbol in executable text run without extra approval,
while commands or script files that would execute symbolized data
require human approval.

\begin{table*}[!t]
\centering
\footnotesize
\setlength{\tabcolsep}{3pt}
\caption{Concrete examples for the two policy types. The data trust
  policy block shows schema rules and user-supplied origin rules. The
  data usage policy block shows executable command
  patterns and command rewriting rules. The dark red text marks
  symbolized fields or command and code text that \HumanShell would
  execute, and a command with no symbol in executable text runs
  without human approval.}
\label{tab:policy-rule-examples}
\begin{tabular}{@{}>{\raggedright\arraybackslash}p{0.25\textwidth}
                  >{\raggedright\arraybackslash}p{0.3\textwidth}
                  >{\centering\arraybackslash}p{0.04\textwidth}
                  >{\raggedright\arraybackslash}p{0.35\textwidth}@{}}
\toprule
\textbf{Policy rule} & \textbf{Tool result or command} & & \textbf{\sys output} \\
\midrule
\multicolumn{4}{@{}l}{\textbf{Data trust policy}
  \textit{(checks tool output)}} \\
\midrule
\textbf{Schema rules}\newline
\textit{Trusted}\newline
\cc{web\_fetch.url}\newline
\textit{Untrusted}\newline
\cc{web\_fetch.text} &
\cc{web\_fetch(url) = \{}\newline
\hspace*{1em}\cc{url=https://reports.example/q1,}\newline
\hspace*{1em}\cc{text="Quarterly report"}\newline
\cc{\}} &
\(\Rightarrow\) &
\cc{web\_fetch(url) = \{}\newline
\hspace*{1em}\cc{url=https://reports.example/q1,}\newline
\hspace*{1em}\cc{text=}\textcolor{Maroon}{\cc{\$web1.text}}\newline
\cc{\}} \\
\addlinespace[0.8ex]
\textbf{Origin rules}\newline
\textit{Trusted}\newline
\cc{api.github.com/*} &
\cc{web\_fetch(https://api.github.com/zen) = \{}\newline
\hspace*{1em}\cc{content="Design for failure."}\newline
\cc{\}} &
\(\Rightarrow\) &
\cc{web\_fetch(https://api.github.com/zen) = \{}\newline
\hspace*{1em}\cc{content="Design for failure."}\newline
\cc{\}} \\
\addlinespace[0.35ex]
\textit{Untrusted}\newline
\cc{imports/*.csv} &
\cc{read(imports/q1.csv) = \{}\newline
\hspace*{1em}\cc{content="name,amount..."}\newline
\cc{\}} &
\(\Rightarrow\) &
\cc{read(imports/q1.csv) = \{}\newline
\hspace*{1em}\cc{content=}\textcolor{Maroon}{\cc{\$read1.content}}\newline
\cc{\}} \\
\addlinespace[0.35ex]
\textit{Untrusted}\newline
\cc{agent:public-chat} &
\cc{session\_history(agent:public-chat) = \{}\newline
\hspace*{1em}\cc{message="check this link"}\newline
\cc{\}} &
\(\Rightarrow\) &
\cc{session\_history(agent:public-chat) = \{}\newline
\hspace*{1em}\cc{message=}\textcolor{Maroon}{\cc{\$session1.message}}\newline
\cc{\}} \\
\addlinespace
\multicolumn{4}{@{}l}{\textbf{Data usage policy}
  \textit{(checks commands sent to \HumanShell)}} \\
\midrule
\textbf{Executable Command Pattern} &
\cc{exec(git status)} &
\(\Rightarrow\) &
No human approval required\\
\addlinespace[0.35ex]
\textcolor{Maroon}{\cc{<symbol>}} &
\cc{exec(}\textcolor{Maroon}{\cc{\$web1.text}}\cc{)} &
\(\Rightarrow\) &
\textcolor{Maroon}{\cc{\$web1.text}} needs human approval for execution \\
\addlinespace[0.35ex]
\cc{python -c }\textcolor{Maroon}{\cc{<symbol>}} &
\cc{exec(python -c }\textcolor{Maroon}{\cc{\$web1.text}}\cc{)} &
\(\Rightarrow\) &
\textcolor{Maroon}{\cc{\$web1.text}} needs human approval for execution \\
\addlinespace
\textbf{Command Rewriting Rule}\newline
\cc{python <file>}\newline
\(\xrightarrow{\textit{transform}}\)\newline
\cc{python -c }\textcolor{Maroon}{\cc{<file content>}} &
\strut\newline
\cc{exec(python fix.py)}\newline
\(\xrightarrow{\textit{transform}}\)\newline
\cc{exec(python -c "...}\textcolor{Maroon}{\cc{\$web1.text}}\cc{...")} &
\strut\newline
\(\Rightarrow\) &
\strut\newline
\textcolor{Maroon}{\cc{\$web1.text}} needs human approval for execution \\
\bottomrule
\end{tabular}
\end{table*}

\PP{PinchBench policy}
Of the 147 PinchBench tasks, 114 read a file as input and 121 write a
file as the goal.
For example, an email reply task reads an inbox under
\cc{emails/*} or \cc{inbox/*} and writes the reply as a file in the
workspace.
For \sys and the Dual LLM pattern, we marked untrusted
input files so that their content is symbolized
(\autoref{tab:pinchbench-untrusted-files}); examples include
\cc{research/*}, \cc{GPT4.pdf}, and \cc{auth.log}, and 42 tasks read at
least one such file.
For Sandboxing, we placed privileged output files, such as email replies
and business outputs, in dedicated directories and denied writes there,
and minimized network access to prevent data leakage and arbitrary
actions (\autoref{tab:pinchbench-output-routes}); the blocked routes
cover 60 tasks.

\end{document}